\pdfoutput=1
\documentclass[12pt]{article}
\usepackage{amsmath,amssymb,latexsym,cite,eucal,amsthm,enumerate}
\usepackage[matrix,arrow,frame,import,curve,color]{xy}
\usepackage{tikz}
\usetikzlibrary{arrows}
\usetikzlibrary{decorations.pathmorphing}
\usetikzlibrary{decorations.markings}
\usetikzlibrary{calc}
\usetikzlibrary{3d}

\usepackage{graphicx}

\newtheorem{theorem}{Theorem}
\newtheorem{prop}[theorem]{Proposition}

\newtheorem{conjecture}[theorem]{Conjecture}

\theoremstyle{definition}

\newif\iffigs\figstrue

\DeclareFontFamily{U}{rsf}{}
\DeclareFontShape{U}{rsf}{m}{n}{
  <5> <6> rsfs5 <7> <8> <9> rsfs7 <10-> rsfs10}{}
\DeclareMathAlphabet\Scr{U}{rsf}{m}{n}

\makeatletter
\newcommand{\miniscule}{\@setfontsize\miniscule{5}{6}}% \tiny: 6/7
\makeatother

\def\O{\Scr{O}}
\def\C{{\mathbb C}}
\def\P{{\mathbb P}}

\def\R{{\mathbb R}}
\def\Z{{\mathbb Z}}

\def\HS{{\mathbb F}}

\def\Hom{\operatorname{Hom}}

\def\Spec{\operatorname{Spec}}
\def\Proj{\operatorname{Proj}}

\def\Conv{\operatorname{Conv}}

\def\Vol{\operatorname{Vol}}

\def\SU{\operatorname{SU}}

\def\Cone{\operatorname{Cone}}

\def\id{{\mathbf{1}}}

\def\CY{Calabi--Yau}
\def\LG{Landau--Ginzburg}

\def\cA{{\Scr A}}
\def\cB{{\Scr B}}

\def\DC{\mathbf{D}^b}

\def\QED{$\quad\blacksquare$}

\def\eqn#1#2{\begin{equation}#2
  \ifx{#1}{}\else\label{#1}\fi\end{equation}}

\begin{document}

\begin{titlepage}
\begin{flushright}
% arXiv:
February 2017
\end{flushright}
\vspace{.5cm}
\begin{center}
\baselineskip=16pt
{\fontfamily{ptm}\selectfont\bfseries\huge
Mirror Symmetry and Discriminants\\[20mm]}
{\bf\large Paul S.~Aspinwall, M.~Ronen Plesser and Kangkang Wang
 } \\[7mm]

{\small

Departments of Mathematics and Physics,\\ 
  Box 90320, \\ Duke University,\\ 
 Durham, NC 27708-0320 \\ \vspace{6pt}

 }

\end{center}

\begin{center}
{\bf Abstract}
\end{center}

We analyze the locus, together with multiplicities, of ``bad''
conformal field theories in the compactified moduli space of $N=(2,2)$
superconformal field theories in the context of the generalization of
the Batyrev mirror construction using the gauged linear
$\sigma$-model. We find this discriminant of singular theories is
described beautifully by the GKZ ``A-determinant'' but only if we use
a noncompact toric \CY\ variety on the A-model side and logarithmic
coordinates on the B-model side. The two are related by ``local''
mirror symmetry.  The corresponding statement for the compact case
requires changing multiplicities in the GKZ determinant. We then
describe a natural structure for monodromies around components of this
discriminant in terms of spherical functors. This can be considered a
categorification of the GKZ $A$-determinant. Each component of the
discriminant is naturally associated with a category of massless
D-branes.

%\vfil\noindent
%\verb$LastChangedRevision: 221 $\\
%\verb$LastChangedDate: 2017-02-15 10:53:38 -0500 (Wed, 15 Feb 2017) $

\end{titlepage}

\vfil\break

%%%%%%%%%%%%%%%%%%%%%%%%%%%%%%%%%%%%%%%%%%%%%%%%%%%%%%%%%%%%%%%%

\section{Introduction}    \label{s:intro}

It is well known that the quintic hypersurface in $\P^4$ is mirror to
an orbifold of another quintic with defining equation
\begin{equation}
  b_1x_1^5 + b_2x_2^5 + b_3x_3^5 + b_4x_4^5 + b_5x_5^5 +
  b_0x_1x_2x_3x_4x_5 = 0,
\end{equation}
where the $b$'s are complex parameters. The quintic has a one
dimensional moduli space of complexified K\"ahler forms, while the
mirror has a one dimensional moduli space of complex structures given
by the parameter $(5\psi)^5 = -b_0^5/(b_1b_2b_3b_4b_5)$. The mirror
quintic becomes singular when $\psi^5=1$ and there is a corresponding
singularity for conformal theories associated with the quintic with
the appropriate K\"ahler form. The limit $\psi\to\infty$ is associated
to the large radius limit of the quintic. For our purposes, this large
radius limit should also be viewed as a kind of singularity in the
natural toric compactification of the
moduli space.  These two points form the ``discriminant''. They
correspond to the locus of ``bad theories''.

More generally, when there are more parameters, the discriminant is a
codimension one subvariety. Typically this discriminant is a very
complicated space. Generally it has more than one component. These
components will intersect each other and may have singularities
themselves. In string theory, the discriminant is associated with
D-branes becoming massless (for suitable normalizations when
considering the large radius limit) and thus often with interesting
physics such as enhanced gauge symmetry or conformal field theories. 

The picture we will see is that each component has a number (or
category) associated to it and this measures the degree of singularity
at a generic point on this component. One may also view this number as
the multiplicity of the component. At singularities of the
discriminant this degree will go higher, but our focus will be on
the generic degree.

So how exactly should we measure the degree of a singularity? As
straight-forward answer might be to compute some partition function
and see the degrees appearing as the order of poles in this
function. We won't do this here, but if we did, there would an
interesting issue of an anomaly as we discuss later. An alternative
approach, which we will use in this paper, is to look at the spectrum
of D-branes and how they become massless at points on the discriminant.

The mathematics literature already contains a very natural answer to
what degrees to attach to the components of the discriminant.  This
comes from the work of Gelfand, Kapranov and Zelevinsky
\cite{GKZ:book}. There the ``GKZ A-determinant'' was defined. In the
case of the quintic, this object is
\begin{equation}
 E = b_1^4b_2^4b_3^4b_4^4b_5^4(b_0^5 + 5^5b_1b_2b_3b_4b_5).
\end{equation}
Clearly the vanishing of $E$ is associated to singularities of the
conformal field theory, but notice the 4th powers appearing.  This
determinant can also be computed for more complicated examples and
again interesting powers appear in many places.  The first purpose of
this paper is to understand how and why these exponents appear using
the language of the gauged linear $\sigma$-model (GLSM). We will tie
these powers to the desired degrees of the components of the
discriminant. However, in order for this to work, we are forced to
consider {\em noncompact\/} mirror symmetry.

Batyrev's mirror symmetry construction \cite{Bat:m} for \CY\
hypersurfaces in toric varieties is naturally generalized by the GLSM
to a whole class of families including complete intersections, but
also models where no geometric large radius limit need exist
\cite{W:phase,Boris:m,AG:gmi}.

We should note that the discriminant plays an important role in mirror
symmetry.  If $X$ and $Y$ are a mirror pair then the complexified
K\"ahler form moduli space of $X$ (suitably corrected) should match
the moduli space of complex structures on $Y$. In particular, the set
of ``bad'' theories due to a K\"ahler degeneration of $X$ should match
the bad theories for $Y$ for a singular complex structure. For mirror
symmetry to work, the discriminants, complete with their
multiplicities should agree between these moduli spaces $X$ and $Y$.
Further, for mirror symmetry to extend to string theory, the massless 
brane spectra should agree.

This complete matching for the case of compact $X$ and $Y$ is actually
quite hard to prove and we will not achieve it fully here. Instead we
focus more on the very closely related question of the associated
noncompact toric ambient \CY's containing $X$ and $Y$, and here we do
find complete agreement. In this case, the discriminant is the GKZ
$A$-determinant and we will show that the discriminants do indeed
match as mirror symmetry implies. However, to get this to work, we
require a change to logarithmic coordinates for $Y$ as seen by Hori
and Vafa \cite{Hori:2000kt} using duality arguments. As we will see,
the GLSM provides a beautiful interpretation for the multiplicities
appearing in the GKZ determinant in terms of noncompact mirror
symmetry.

The general idea is that for each component of the discriminant, the
GLSM can be broken up into a Higgs factor and a Coulomb
factor. The Coulomb factor describes the breakdown of the conformal
field theory in terms of some scalar fields becoming
noncompact. Meanwhile, the multiplicity of this component in the
discriminant is encoded in the Higgs factor.

It should be emphasized that the notion of multiplicity in the
noncompact case seems to be very natural. As well as the rank of the
charge lattice of massless D-branes, it has an interpretation in terms
of the Witten index, or orbifold Euler characteristic of the Higgs
theory. This coincides with the rank of the cohomology ring and thus
the topological K-theory($K_0$) of the model in question. Furthermore,
the toric geometry gives a coincidence between the cohomology ring and
the Chow ring. This implies an equality between the rank of
topological $K_0$ and algebraic $K_0$.

Being the rank of the group $K_0$, the multiplicities are naturally
``categorified''.  We tie this in with the a natural description of
monodromy around the discriminant. As we said above, the
categorification in question is the ``category of massless D-branes''
as explored in \cite{AdAs:masscat}. The obvious thing to do is to
identify this category of massless D-branes as precisely the category
of D-branes of the Higgs theory. We will see this works in many cases.

The description becomes more awkward when we pass to the compact
case. However, we note that matching the location of the discriminant
and monodromy between the two sides may be viewed as a route to
provide yet another ``proof'' of mirror symmetry. This might be
regarded as somewhat easier than going via a computation
Gromov--Witten indices requiring a knowledge of period
integrals. However, they should, of course, be related by the
Hilbert--Riemann correspondence.

We begin in section \ref{sec:GLSM} by reviewing the general setup for
the abelian GLSM. Then in section \ref{s:Bside} we discuss how the
discriminant appears in terms of the complex structure model space,
i.e., the B-model moduli space. This ties the discriminant analysis of
the GLSM into the work of GKZ, but we find that there they do {\em
  not\/} fully coincide. In section \ref{s:Aside} we then analyze how
the A-model becomes singular. Each component of the discriminant is
then described by a split of GLSM into a Higgs part and Coulomb part,
where the Higgs part determines the multiplicity. We also see how to
resolve the mismatch found in section \ref{s:Bside} by going to the
noncompact model.

In section \ref{s:mon} we recast the story in terms of D-brane
categories. This turns out to give a very natural picture of the
components of the discriminant. The D-brane category associated to
each component is the source of a functor giving the generic
monodromy around the discriminant in terms of a ``spherical
twist''. Finally we give concluding remarks in section \ref{s:conc}.

%%%%%%%%%%%%%%%%%%%%%%%%%%%%%%%%%%%%%%%%%%%%%%%%%%%%%%%%%%%%%%%%%%%

\section{General Toric Mirror Symmetry} \label{sec:GLSM}

Here we review the set-up of toric data to define the GLSM and the
resulting toric mirror symmetry statement. This is based on Batyrev's
hypersurface construction \cite{Bat:m} but then generalized to include
all abelian GLSM's as in \cite{W:phase}. We refer to
\cite{AP:genmir} for details. The data is as follows:

\begin{itemize}
\item On the ``A side'' we have a $d$-dimensional lattice $N$, a set
  of $n$ points $\cA$ in $N$, $n$ complex parameters
  $(a_1,\ldots, a_n)\in\C^n$ and a vector $\nu\in N$.
\item On the ``B side'' we have a lattice $M=N^\vee$, a set of $m$ points
$\cB$ in $M$, $m$ complex parameters
$(b_1,\ldots, b_m)\in\C^m$ and a vector $\mu\in M$.
\item The pointset $\cA$ lies in a hyperplane
  $\langle\mu,\alpha\rangle=1$, for all $\alpha\in\cA$ and the
  pointset $\cB$ lies in a hyperplane $\langle\beta,\nu\rangle=1$,
  for all $\beta\in\cB$.
\item We define the superpotential in $\C[\cA] = \C[x_1,\ldots,x_n]$,
\begin{equation}
  W = \sum_{\beta\in\cB} b_\beta x^{\beta}, \quad
     x^\beta = \prod_{\alpha\in\cA}x_\alpha^{\langle\beta,\alpha\rangle}.
\end{equation}
\item We define a height function $\zeta:\cA\to\R$ given by
  $\log|a_\alpha|$. For generic values of $a_\alpha$ this gives a
  triangulation of $\cA$ (or ``phase'') and an associated fan $\Sigma$
  over this triangulation. The particular triangulation we use is the
  one consistent with a convex piece-wise linear function taking the
  values of the height function at each point in $\cA$.
\end{itemize}

The fan $\Sigma$ produces a toric variety $Z_\Sigma$ of dimension $d$
in the usual way. $Z_\Sigma$ is noncompact and \CY\ because of the
hyperplane condition. When $Z_\Sigma$ is smooth, the height function
$\zeta$ can be used to produce a K\"ahler metric on $Z_\Sigma$ via
symplectic reduction \cite{AGM:II}. 

We then define the classical target space $X_\Sigma\subset Z_\Sigma$
as the critical point set of the superpotential $W$. For a suitable
choice of phase, $X_\Sigma$ may be a compact \CY\ manifold of dimension
$d-2\langle\mu,\nu\rangle$. The metric on $Z_\Sigma$ induces a metric
on $X_\Sigma$ but this is not desired one. The true metric is realized
through IR flow in the GLSM to the conformal field theory.

More generally $X_\Sigma$ will be
associated with an $N=(2,2)$ superconformal field theory, and
$X_\Sigma$ may have the interpretation of an orbifold, Landau--Ginzburg
theory, some LG-fibration hybrid model, etc.

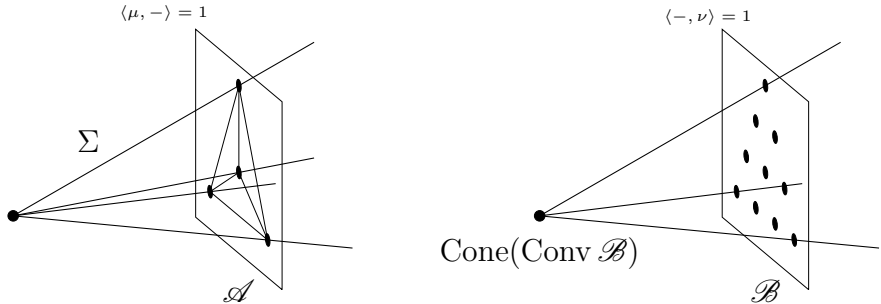
\begin{figure}
\begin{center}
\tikzset{xzplane/.style={canvas is xz plane at y=#1,very thin}}
\tikzset{yzplane/.style={canvas is yz plane at x=#1,very thin}}
\tikzset{xyplane/.style={canvas is xy plane at z=#1,very thin}}
\begin{tikzpicture}[x={(140:0.5cm)},y={(0cm,1cm)},
    z={(1cm,0cm)}]
\draw[xyplane=3] (-1.5,-0.5) -- (1.5,-0.5) -- (1.5,2) -- (-1.5,2) -- (-1.5,-0.5);
\draw[xyplane=3] (-1,0) -- (1,0) -- (0,{sqrt(3)}) -- (-1,0);
\draw[xyplane=3] (0,{sqrt(3)}) -- (0,{1/sqrt(3)}) -- (1,0);
\draw[xyplane=3] (0,{1/sqrt(3)}) -- (-1,0);
\filldraw[xyplane=3] (-1,0) circle (0.08);
\filldraw[xyplane=3] (1,0) circle (0.08);
\filldraw[xyplane=3] (0,{sqrt(3)}) circle (0.08);
\filldraw[xyplane=3] (0,{1/sqrt(3)}) circle (0.08);
\draw (0,0,0) -- ({-4/3},0,4);
\draw (0,0,0) -- ({4/3},0,4);
\draw (0,0,0) -- (0,{4/3*sqrt(3)},4);
\draw (0,0,0) -- (0,{4/3/sqrt(3)},4);
\draw (0,-1,3) node {$\cA$};
\draw (0,2.7,2) node {\tiny $\langle\mu,-\rangle=1$};
\filldraw[yzplane=0] (0,0) circle (0.07);
\draw (0,1,1) node {$\Sigma$};
\draw[xyplane=10] (-1.5,-0.5) -- (1.5,-0.5) -- (1.5,2) -- (-1.5,2) -- (-1.5,-0.5);
\filldraw[xyplane=10] (-1,0) circle (0.08);
\filldraw[xyplane=10] ({-1/3},0) circle (0.08);
\filldraw[xyplane=10] ({1/3},0) circle (0.08);
\filldraw[xyplane=10] (1,0) circle (0.08);
\filldraw[xyplane=10] (0,{sqrt(3)}) circle (0.08);
\filldraw[xyplane=10] ({1/3},{2/3*sqrt(3)}) circle (0.08);
\filldraw[xyplane=10] ({2/3},{1/3*sqrt(3)}) circle (0.08);
\filldraw[xyplane=10] ({-1/3},{2/3*sqrt(3)}) circle (0.08);
\filldraw[xyplane=10] ({-2/3},{1/3*sqrt(3)}) circle (0.08);
\filldraw[xyplane=10] (0,{1/sqrt(3)}) circle (0.08);
\draw (0,0,7) -- ({-4/3},0,11);
\draw (0,0,7) -- ({4/3},0,11);
\draw (0,0,7) -- (0,{4/3*sqrt(3)},11);
\draw (0,-1,10) node {$\cB$};
\draw (2,2,10) node {\tiny $\langle-,\nu\rangle=1$};
\filldraw[yzplane=0] (0,7) circle (0.07);
\draw (0,-0.5,7) node {$\Cone(\Conv\cB)$};
\end{tikzpicture}
\end{center}
\caption{The data used to define a GLSM.} \label{fig:data}
\end{figure}

The topological A-model \cite{W:AB} associated to this depends only
the $a_\alpha$ parameters, which determine the (complexified) K\"ahler
form. The topological B-model depends only on the $b_\beta$
parameters, which determine the complex structure when $X_\Sigma$ is a
smooth \CY. Mirror symmetry is an exchange of the A-side and
B-side. That is, if $Y$ is mirror to $X$ then $Y$ has its A-model data
given by $b_\beta$ and the B-model data given $a_\alpha$. Thus, we not
only identify $X$ and $Y$ as mirrors. but we also {\em globally\/}
identify their parameter spaces. Note that $(\C^*)^d$ acts on both set
of parameters $a_\alpha$ and $b_\beta$ and that the resulting GLSM
appears invariant under this action. Thus one might divide out by this
action to obtain a more effective parametrization. We will find it
useful to keep this apparently redundant parametrization most of the
time. There could also be an anomaly associated with this $(\C^*)^d$
action as we will see in section \ref{sss:anomaly}.

It is worth emphasizing that we are using the $a_\alpha$'s to
parametrize the K\"ahler moduli space and not the actual K\"ahler
form cohomology class. The latter requires us to pass to ``flat
coordinates'' by solving Picard--Fuch's equations. We completely evade
this process by sticking with the ``algebraic coordinates'' given by
$a_\alpha$. One can also argue that these algebraic coordinates are
more naturally associated both with the toric structure and with the
superconformal theory moduli space.

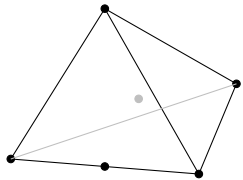
\begin{figure}
\begin{center}
\begin{tikzpicture}
\filldraw(0,0) circle (0.05);
\filldraw(2.5,-0.2) circle (0.05);
\filldraw(1.25,2) circle (0.05);
\filldraw(3,1) circle (0.05);
\filldraw(1.25,-0.1) circle (0.05);
\filldraw[lightgray] (1.7,0.8) circle (0.05);
\draw (1.25,2) -- (0,0) -- (2.5,-0.2) -- (1.25,2) -- (3,1) -- (2.5,-0.2); 
\draw[lightgray] (0,0) -- (3,1);
\end{tikzpicture}
\end{center}
\caption{The pointset $\cA$ for the octic
hypersurface in $\P^4_{2,2,2,1,1}$ (with one dimension missing).}   \label{fig:octic}
\end{figure}

Let us fix an example to clarify the procedure. We use the ``octic
hypersurface'' in $\P^4_{2,2,2,1,1}$ as analyzed in many places
including \cite{CDFKM:I}. Here $\cA$ consists of 7 points, and
$d=5$. Five points lie at the vertices of a 4-simplex, one point lies
in the proper interior of this simplex and the last point lies at the
midpoint of a one-dimensional edge. We show this reduced by one
dimension in figure \ref{fig:octic}. There are 4 triangulations of
$\cA$. The unique triangulation involving all 7 points gives
$Z_\Sigma$ as a line bundle over the resolution of $\P^4_{2,2,2,1,1}$,
and $X_\Sigma$ as the smooth \CY\ hypersurface in this resolved
weighted projective space. The pointset $\cB$ has 105 points
corresponding to the 105 monomials of degree 8 on $\P^4_{2,2,2,1,1}$.

Going to the mirror by exchanging $\cA$ and $\cB$, the pointset $\cB$
now has 7 elements. Thus the superpotential for the mirror $Y_\Sigma$
has 7 monomials in 105 homogeneous coordinates. If we only include the 5
coordinates $x_\alpha$ associated to the vertices of $\cA$ (as we would for the
\LG\ phase) the superpotential can be written
\begin{equation}
W = b_1x_1^4 + b_2x_2^4 + b_3x_3^4 + b_4x_4^8 + b_5x_5^8
  + b_6x_4^4x_5^4 + b_7x_1x_2x_3x_4x_5. \label{eq:Woct}
\end{equation}

%%%%%%%%%%%%%%%%%%%%%%%%%%%%%%%%%%%%%%%%%%%%%%%%%%%%%%%%%%%%%%%%%%%

\section{The B-Side Discriminant}  \label{s:Bside}

\subsection{$\Delta_\cB$} \label{s:GKZdisc}

We now review the motivation and construction of the
``$A$-determinant'' of \cite{GKZ:book}. To fit in with the notation
used in this paper, we will need to call it the $\cB$-determinant for now. We
begin with the definition of an associated object $\Delta_\cB$.

Assume we are given a set of points $\cB$ in a hyperplane in
$M\cong\Z^d$. Let $\{e_1,\ldots,e_d\}$ be a basis of the dual lattice
$N=M^\vee$. We can define a Laurent polynomial in $\C[y_1,\ldots,y_d]$.
\begin{equation}
W = \sum_{\beta\in\cB} b_\beta \prod_{i=1}^d y_i^{
 \langle b_\beta,e_i\rangle},
\end{equation}
for complex parameters $b_\beta$.
We now might ask if the hypersurface $W=0$ in $\C^d$ is singular. That
is, is there a solution of
\begin{equation}
 W = \frac{\partial W}{\partial y_1} =\ldots=
   \frac{\partial W}{\partial y_d} = 0,  \label{eq:d1}
\end{equation}
for some $y_i$? Since we might have negative powers of $y_i$, we
should restrict the question to all the $y_i$'s being nonzero, in
order to have a well-posed question. There will be a subset of
parameters $b_\beta$ for which we can find a solution of
(\ref{eq:d1}). Assuming this pointset is codimension one, this gives an
irreducible function $\Delta_{\cB}$ of $b_\beta$, such that the
(Zariski closure of) the above subset of parameters is given by
$\ \Delta_{\cB}=0$.

GKZ refer to $\Delta_{\cB}$ as the ``discriminant'' of $W$. Indeed one
usually associates the vanishing of the discriminant to
singularities. In this paper we will avoid using the word
``discriminant'' for $\Delta_{\cB}$. This is because $\Delta_{\cB}=0$
represents only some of the possibilities for singularities. In particular,
the constraint that all $y_i$'s be nonzero is too strong. We will use
the term discriminant to refer to the {\em complete\/} locus of ``bad'' GLSM's.
 
One can compute $\Delta_{\cB}$ most efficiently using the ``Horn
uniformization'' trick \cite{MR1109634,GKZ:book}. Let $B$ be the
$d\times m$ matrix given by the coordinates of the points
$\beta\in\cB$, and let $S$ be the $r\times m$ matrix given as the
kernel of $B^t$ over the integers, where $r=m-d$. Let $s_{i\beta}$ be the integral
entries of $S$. We then have $(\C^*)^d$-invariant coordinates
\begin{equation}
  z_i = \prod_{\beta\in\cB} b_\beta^{s_{i\beta}}, \label{eq:zinv}
\end{equation}
parametrizing the complex structure moduli space. $\Delta_{\cB}$ is
then given in parametrized form
\begin{equation}
  z_i = \prod_{\beta\in\cB} \Biggl(\sum_{j=1}^{r}s_{j\beta}\lambda_j
             \Biggr)^{s_{i\beta}},  \label{eq:Horn}
\end{equation}
where $[\lambda_1,\lambda_2,\ldots,\lambda_{r}]\in\P^{r-1}$.
That is, we may find $\Delta_{\cB}$ by eliminating
$\lambda_1,\lambda_2,\ldots,\lambda_{r}$ from (\ref{eq:Horn}).

In the case that $r$ is around 4 or less, a computer algebra
package may perform the elimination required, but for larger values,
$\Delta_{\cB}$  has many many terms and becomes impractical to
compute.

The elements of $\cB$ may be viewed as monomials in the homogeneous
coordinate ring, with coordinates corresponding to points in
$\cA$. With this interpretation, let $Z_0=\Spec\C[\cB]$. Note that if
$\Cone\Conv(\cA)$ is dual to $\Cone\Conv(\cB)$, i.e., we are in the
reflexive case, then $Z_0$ is the affine toric variety corresponding
to the cone over $\Conv\cA$. That is, $Z_0$ corresponds to the
``non-simplicial-decomposition'' of $\cA$. $Z_0$ is a {\em phase\/}
$Z_\Sigma$ if and only if $\Conv\cA$ is itself a simplex.

Let $\Omega^\bullet_{Z_0}$ denote the ring of differential forms on
$Z_0$. Note that, in addition to the usual degree of a differential
form, the $R$-charge vector $\nu$ induces another degree on this ring,
where every monomial in $\cB$ has $R$-charge 1.
One can then consider the Koszul complex:
% Cor 4.2 pg 72 of GKZ
\begin{equation}
\xymatrix@1{
\Omega^0_{Z_0}\ar[r]^-{\wedge dW}&\Omega^1_{Z_0}\ar[r]^-{\wedge
  dW}&\ldots}
\label{eq:dWKoz}
\end{equation}
This complex will be exact if and only if $dW$ is nowhere vanishing on
$Z_0$. We may turn this into a complex of vector spaces by taking
global sections and restricting to a particular
$R$-charge.\footnote{For computations an arbitrarily large $R$-charge is chosen so that
  higher cohomologies vanish. That is, we have a {\em stable
    twisting\/} \cite{GKZ:book}.} 

Using a method due to Cayley, one may compute a ``determinant'' of a
complex of vector spaces which vanishes precisely when the complex
fails to be exact. GKZ prove that this determinant correctly computes
$\Delta_{\cB}$ but only in the rather restrictive case that
$\Proj\C[\cB]$ is smooth. As well as this rather restrictive
condition, which is violated in almost any case we would like to
consider, the description of differential forms on a toric variety is a
little unnatural when it comes to computations. We discuss the
resolution of this problem shortly.

Let us consider the singularities of the B-model in the case that we
have a geometrical interpretation. The simplest case is when
$X_\Sigma$ is a hypersurface in a toric variety. Then $W$ is of the
form
\begin{equation}
  W = x_0f(x_1,x_2,\ldots).
\end{equation}
A compact toric variety $V\subset Z$ is given by $x_0=0$, and the hypersurface
$X\subset V$ by $f=0$. As discussed in the next section, the theory is singular
if $x_0$ is not forced to be zero by the constraint $dW=0$, since that
would give us a noncompact target space. Since
$dW = fdx_0 + x_0df$, the theory is singular when $f=df=0$ has a
solution (for suitably nonzero $x_\alpha$). This is the familiar
condition for a hypersurface singularity. Restricting to the
hypersurface $f=0$, the degree of $dW$ is the same as the degree of
$df$. In the case of isolated singularities this is equal to the {\em
  Milnor Number\/} of the singularity. A generic deformation of the
singularity hypersurface will produce a smooth hypersurface with a
bouquet of spheres growing out of the singularity. The number of
spheres is the Milnor number.

So in this special case we see that the degree of vanishing of $dW$
gives the Milnor number which measures the complexity of the
singularity. In the most general case, there is not a geometrical
interpretation so we cannot always have this Milnor number
interpretation.

\subsection{The GKZ $\cB$-Determinant} \label{s:GKZ}

\subsubsection{Definition}

In order to analyze some general properties of the discriminant, GKZ
introduced the $\cB$-determinant\footnote{$E_\cB$ is only defined up
  to a sign.} $E_{\cB}$ \cite{GKZ:book}. This has,
in general, even more terms that $\Delta_{\cB}$ and is thus even
more impractical to completely compute for generic examples.
But in some ways it has a nicer structure. In particular it is easy to
compute the coefficients of some of the monomials in $E_{\cB}$ which
have a correspondence to vertices of the convex hull of the Newton
polytope. More importantly for us, the $\cB$-determinant will be a
very natural object to study in the context of the GLSM.

Let $Z_1\subset Z_0$ be the union of all the toric coordinate
hyperplanes. That is, $Z_1$ is given by a divisor where any
homogeneous coordinate vanishes. Let $\Omega^\bullet_{Z_0}(\log Z_1)$
be the ring of {\em logarithmic\/} differential forms on $Z_0$. Such
differential forms are allowed simple poles along $Z_1$. That is, if
$y_i$ are affine coordinates in a patch where $Z_1$ corresponds to the
coordinate hyperplanes, then we consider forms with terms like
\begin{equation}
  f(y_1,y_2,\ldots)\,\frac{dy_i}{y_i}\wedge\frac{dy_j}{y_j}\ldots
\end{equation}
If we put $Y_j=\log y_j$, then such a form becomes holomorphic:
\begin{equation}
  f(e^{Y_1},e^{Y_2},\ldots)\,dY_i\wedge dY_j\ldots,
\end{equation}
hence the notation $\Omega^\bullet_{Z_0}(\log Z_1)$.

Logarithmic differential forms are much more natural in toric geometry
than normal differential forms since they are invariant under the
torus action. A basis for sections of the cotangent sheaf
of $T_N\cong N\otimes_\Z\C^*\cong(\C^*)^d$ is given by
\begin{equation}
\frac{dy_1}{y_1},\frac{dy_2}{y_2},\ldots,\frac{dy_d}{y_d}.
\end{equation}
(See exercise 8.1.1 of \cite{CLS:ToricVar}.)  Indeed, the best way to
construct $\Omega^\bullet_{Z_0}$ is to first construct
$\Omega^\bullet_{Z_0}(\log Z_1)$ and then impose the condition that
there are no poles along $Z_1$. This restriction is described in
definition 2.3 in chapter 9 of \cite{GKZ:book}.

It is combinatorially more natural than (\ref{eq:dWKoz}), therefore,
to consider the Koszul complex
\begin{equation}
\xymatrix@1@C=10mm{
\Omega^0_{Z_0}(\log Z_1)\ar[r]^-{\wedge dW}&
\Omega^1_{Z_0}(\log Z_1)\ar[r]^-{\wedge dW}&\ldots}
  \label{eq:logdW}
\end{equation}
Taking global sections of the terms in this complex and fixing an
R-charge we may again compute the Cayley determinant. This is the GKZ
$\cB$-determinant $E_{\cB}$.

\subsubsection{Relation to Discriminants}

One of the main results of GKZ \cite{GKZ:book} is that there is a
relationship between $E_{\cB}$ and $\Delta_{\cB}$:
\begin{equation}
  E_{\cB} = \prod_\Gamma
  (\Delta_{\cB\cap\Gamma})^{u(\Gamma)i(\Gamma)},
     \label{eq:Adet}
\end{equation}
where the product is taken over all the faces $\Gamma$ of
$\Conv\cB$. The faces are of any dimension and include the whole of
$\Conv\cB$. The $u(\Gamma)$'s and $i(\Gamma)$'s are nonnegative
integers we will describe shortly.

Note that in the previous section we only really defined
$\Delta_{\cB\cap\Gamma}$ up to an overall constant. The GKZ definition
of the $\cB$-determinant as a function of the $b_\beta$'s completely
fixes everything up to an overall sign. One manifestation of this is
that, while the Horn uniformization of equations (\ref{eq:zinv}) and
(\ref{eq:Horn}) imply the corresponding $\Delta_{\cB\cap\Gamma}$'s are
$(\C^*)^d$-invariant, the $\cB$-determinant turns out {\em not\/} to
be $(\C^*)^d$-invariant. However, this $(\C^*)^d$ dependence is purely
an overall factor. We'll discuss this more below, but we note now that
the divisor determined by $E_\cB=0$ {\em is\/} $(\C^*)^d$-invariant.

We need to be a little careful about how exactly we form the product
(\ref{eq:Adet}). If $\Gamma$ is a face of dimension $>0$ and
$\cB\cap\Gamma$ consists purely of points at the vertices of a simplex
then there is no affine relation between them. A such, when we form
the Horn uniformization, $r=0$. The result is that
$\Delta_{\cB\cap\Gamma}=1$. So we ignore faces with no affine
relation. An exception is made for vertices. In that case we set
\begin{equation}
  \Delta_{\{\beta\}} = b_\beta. \label{eq:vetx}
\end{equation}

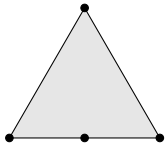
\begin{figure}
\begin{center}
\begin{tikzpicture}
\draw[fill=gray!20!white] (0,0) -- (2,0) -- (1,1.73) -- (0,0);
\filldraw (0,0) circle (0.05);
\filldraw (1,0) circle (0.05);
\filldraw (2,0) circle (0.05);
\filldraw (1,1.73) circle (0.05);
\end{tikzpicture}
\end{center}
\caption{A redundant face not included in the GKZ determinant.}
    \label{fig:redun}
\end{figure}

Faces $\Gamma$ are also excluded from the product in (\ref{eq:Adet})
if they are redundant in the sense that any affine relation between
points in $\cB\cap\Gamma$ can be expressed in terms of affine
relations between points in $\cB\cap\Xi$, where $\Xi$ is a proper
subface of $\Gamma$. Such faces produce a factor $\Delta_\Gamma$ that
coincides with a factor given by a lower-dimensional face. Including
such redundant faces would give incorrect multiplicities.  A redundant
face is shown in figure \ref{fig:redun}.

The quantity $i(\Gamma)$ is defined as follows. The vectors from the
origin to points in the face $\Gamma$ span a linear subspace
$\Gamma_\R\subset M_\R$. The points in $\cB\cap\Gamma$ span a
sublattice $\Gamma_\Z\subset M$. $i(\Gamma)$ is defined as the index
$[\Gamma_\R\cap M:\Gamma_\Z]$. In simple cases $i(\Gamma)=1$ for all
faces. This is true in all the examples we will consider. To simplify
discussion we will ignore this quantity even though it can be properly
taken into account.

The quantity $u(\Gamma)$ is more interesting and is central to this
paper. A set of points in a lattice like we consider in this paper
generate a semigroup $S$ under addition. The quotient $S/\Gamma$ is
also a semigroup and can be thought of as a set of points in the
quotient lattice $M/\Gamma_\Z$.  Let $(S/\Gamma)_+$ denote the set of
points (or semigroup) with the origin removed.

Let $\Cone(S/\Gamma)_+$ be the cone generated by rays from the origin
to any element of $(S/\Gamma)_+$. Note that this is clearly a pointed
cone (i.e., contains no linear subspace), since $\Gamma$ is a face.
Let $\Conv(S/\Gamma)_+$ be the convex hull as usual. Then
\begin{equation}
  u(\Gamma) = \Vol\Bigl(\Cone(S/\Gamma)_+ - \Conv(S/\Gamma)_+\Bigr),
\end{equation}
where the volume is taken relative to the lattice $M/\Gamma_\Z$ using
the usual normalization that a basic simplex is volume 1.

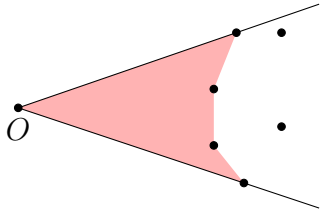
\begin{figure}
\begin{center}
\begin{tikzpicture}[scale=1.0]
\fill[fill=red!30,line width=0pt] (-2,1) -- (0.9,2) -- (0.6,1.25) 
     -- (0.6,0.5) -- (1,0) -- (-2,1);
\filldraw (-2,1) circle (0.05);
\filldraw (1,0) circle (0.05);
\filldraw (0.6,0.5) circle (0.05);
\filldraw (0.6,1.25) circle (0.05);
\filldraw (0.9,2) circle (0.05);
\filldraw (1.5,2) circle (0.05);
\filldraw (1.5,0.75) circle (0.05);
\draw (-2,1) -- (2,-1/3);
\draw (-2,1) -- (2,1+4/2.9);
\draw (-2,1) node[anchor=north] {$O$};
\end{tikzpicture}
\end{center}
\caption{The shape in the computation of $u(\Gamma)$.} \label{fig:vol}
\end{figure}
We show this volume in figure \ref{fig:vol}. Note that the base of the
``pyramid'' will be concave, as shown in the figure, or flat.

Since the polynomials $\Delta_{\cB\cap\Gamma}$ are irreducible, the
expression (\ref{eq:Adet}) is precisely the factorization of $E_\cB$
and the numbers $u(\Gamma)$ give the multiplicity of each irreducible
component.

A special mention should be made for the case where $\Gamma$ is the
whole of $\Conv\cB$. Assuming this face is not redundant, we set
$u(\Gamma)=1$. This factor $\Delta_\cB$ is called the ``primary
component''. In the Batyrev case, where $\langle\mu,\nu\rangle=1$,
there is a point in the proper interior of the $\Conv\cB$ and so
this maximal $\Gamma$ cannot be redundant. Thus, in this case the
primary component always appears as a factor of $E_\cB$ with
multiplicity one. In general,
there are cases where there the primary component does not appear.

The vertices of $\Conv\cB$ giving the monomial contributions
(\ref{eq:vetx}) are associated to the phase picture of the GLSM. When
we view the moduli space as a toric variety in the usual way, some
(but not all) of the toric divisors are associated with these vertices.

Returning to our octic (mirror) example, let $\cB$ be given by the
seven points in figure \ref{fig:octic} and (\ref{eq:Woct}). Here we
have a primary component given by the full simplex. We also have a
contribution to $E_\cB$ coming from the 3 points along an edge, which
comes with multiplicity 3; as well as contributions from the 5
vertices. The result is\footnote{We will always ignore the overall
  sign ambiguity in $E_\cB$.}
\begin{equation}
E_\cB = b_1^6b_2^6b_3^6b_4^4b_5^4\,
(b_6^2-4b_4b_5)^3\,
(2^{18}b_1^2b_2^2b_3^2b_4b_5 -
2^{16}b_1^2b_2^2b_3^2b_6^2+ 2^9b_1b_2b_3b_6b_7^4 - b_7^8).
\end{equation}
The torus $T_N=(\C^*)^d$ acts on $b_1,\ldots,b_9$ to leave 2 invariant
coordinates on the moduli space 
\begin{equation}
x = \frac{b_1b_2b_3b_6}{b_7^4},\quad
y = \frac{b_4b_5}{b_6^2}.
\end{equation}
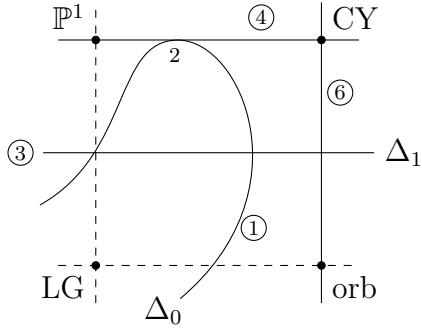
\begin{figure}
\begin{center}
\begin{tikzpicture}
\draw[dashed] (-0.5,0) -- (3.5,0);
\draw[dashed] (0,-0.5) -- (0,3.5);
\draw (-0.5,3) -- (3.5,3);
\draw (3,-0.5) -- (3,3.5);
\filldraw(0,0) circle (0.05) node[anchor=north east] {LG};
\filldraw(3,0) circle (0.05) node[anchor=north west] {orb};
\filldraw(0,3) circle (0.05) node[anchor=south east] {$\P^1$};
\filldraw(3,3) circle (0.05) node[anchor=south west] {CY};
\draw (-0.7,1.5) -- (3.7,1.5) node[anchor=west] {$\Delta_1$};
\begin{scope}[xshift=-68,yshift=-87,scale=0.103]                   ]
\draw (16.0,37.5) .. controls (27.8,44.1) and (25.5,58.8)
.. (33.8,58.8) .. controls (42.1, 58.8) and (50.4,39.1)
.. (34.1, 25.4);
\end{scope}
\draw (0.9,-0.6) node {$\Delta_0$};
\draw (2.1,0.5) node[circle,inner sep=1pt,draw] {\scriptsize 1};
\draw (-1.0,1.5) node[circle,inner sep=1pt,draw] {\scriptsize 3};
\draw (3.25,2.3) node[circle,inner sep=1pt,draw] {\scriptsize 6};
\draw (2.2,3.3) node[circle,inner sep=1pt,draw] {\scriptsize 4};
\draw (1.05,2.8) node {\scriptsize 2};
\end{tikzpicture}
\end{center}
\caption{The discriminant for the octic example.}  \label{fig:octD}
\end{figure}
This gives a discriminant as shown in figure \ref{fig:octD} (drawing
complex dimensions as real). The 4 toric points are associated with
phase limits of the mirror as studied in \cite{CDFKM:I} and we label
these points accordingly in the figure. The numbers in circles show
the multiplicity. Note that two of the four toric divisors ($b_6=0$
and $b_7=0$) are {\em not\/} in the discriminant and are shown as
dashed lines in the figure.

\subsubsection{The ``anomalous'' degree of $E_\cB$}  \label{sss:anomaly}

The polynomial $W$ has $M$ monomials in the $y$'s with coefficients
$b_\beta$.  The $(\C^*)^d$ action on the $y$'s thus induces a
$(\C^*)^d$-action on the $b_\beta$'s and therefore $E_\beta$. The
determinant $E_\cB$ is therefore associated to a character of
$(\C^*)^d$, which is given by a vector $D\in M$.

If $E_\beta$ were to be computed directly as a quantity in the GLSM,
one would expect $D$ to be zero. This is because the $(\C^*)^d$
rescalings are simple field redefinitions and should thus have no
effect classically. A nonzero $D$ would represent an anomaly. It is of
central interest, therefore, to compute $D$.

$D$ may be calculated directly from the definition of $E_\beta$. The
result is as follows. Let $S$ be the semigroup generated by $\cB$
again. Let $S_l$ denote the elements $s\in S$ with $R$-charge $l$, i.e.,
$l= \langle s,\nu\rangle$.

The result is as follows. Identifying $s\in S$ with a vector in the
lattice $M$, we define
\begin{equation}
  D_k = \sum_{l=0}^d (-1)^{d+l}\binom dl\sum_{s\in S_{k+l}} s.
\end{equation}
It follows from the definition of $E_\beta$ in chapter 10, section 1A
of \cite{GKZ:book} that
\begin{equation}
  D = \lim_{k \to\infty} D_k.
\end{equation}
That is, there is some number $K$ such that $D_k=D$ for all $k\geq K$.

Define 
\begin{equation}
\bar s_{l} = \frac{\sum_{s\in S_l}s}{l|S_l|}.
\end{equation}
Then $\bar s_l$ is a vector pointing in the direction of the center of
mass of $S_l$ normalized to have R-charge 1. This gives
\begin{equation}
  D = \lim_{k\to\infty}\sum_{l=0}^d(-1)^{d+l}\binom
  dl|S_{k+l}|(k+l)\bar s_{k+l}. \label{eq:D1}
\end{equation}
The quantity $|S_l|$ is given by the Ehrhart polynomial (see, for
example, \cite{MS:combcomm}). Comparing coefficients of $x^m$ in 
the series expansion of
\begin{equation}
  e^{kx}(e^x-1)^d = \sum_{l=0}^d(-1)^{d+l}\binom dle^{(l+k)x},
\end{equation}
we see
\begin{equation}
  \sum_{l=0}^d(-1)^{d+l}\binom dl(l+k)^m=\begin{cases}
   0 &\quad\text{for } m<d\\
   d! &\quad\text{for }m=d\end{cases}
\end{equation}

The sum in (\ref{eq:D1}) therefore picks out the leading term of the
Ehrhart polynomial, which is given by $(d-1)!$ times the volume of the
pointset $B$.\footnote{Assuming the points in $B$ generate $M$ as an
  abelian group.} This yields
\begin{equation}
  D = d\Vol(B)\bar s,
\end{equation}
where $\bar s$ denotes the limit of $\bar s_l$ as $l\to\infty$, i.e.,
the normalized ``center of mass'' of the cone.

So, in particular, $D$ is not zero. $D$ points in the direction of the
center of mass of the cone over $B$. Note that this direction is not, in
general, the same direction as the vector $\mu\in M$ introduced earlier.

\subsection{Locating Bad Theories}  \label{ss:bad}

Now we will try to relate the GKZ $\cB$-determinant to the
discriminant of bad theories in the moduli space of complex
structures. This was analyzed in detail in \cite{AP:genmir} so we will
review it only briefly here. We analyze the conditions for the {\em
  compact\/} model $X_\Sigma$ to be nonsingular but we will find it
does not reproduce $E_\cB$.

The criterion for a GLSM being singular is that the classical vacuum
is not compact. In \cite{AP:genmir} it was shown that this can be
determined by whether the $\C^*$ orbit associated with the
vector $\nu\in N$, i.e., the R-charge, is compact in $Z_0$. 

The key short exact sequence for the toric construction is
\begin{equation}
\xymatrix@1{0\ar[r]&M\ar[r]^{A^t}&\Z^{\cA}\ar[r]^Q&\widehat G\ar[r]&0},
\end{equation}
where $A$ is the $d\times n$ matrix representing the coordinates of
$\alpha$ and $Q$ is the $r\times n$ matrix so that $Q^t$ is the kernel
of $A$, where $r=n-d$. $\widehat G$ is the dual group $\Hom_Z(G,\C^*)$,
where $G$ is the gauge symmetry of the GLSM.  $Q$ is the matrix of
``charges'' of the homogeneous coordinates $x_\alpha$ under the
$(\C^*)^r\subset G$ quotient action. We can construct affine
coordinates $y_i$ as in the previous section from
\begin{equation}
  y_i = \prod_{\alpha\in\cA} x_\alpha^{A_{i\alpha}},
\end{equation}
so that
\begin{equation}
  y_i\frac{\partial W}{\partial y_i} = \sum_{\alpha\in\cA}
     A_{i\alpha}x_\alpha\frac{\partial W}{\partial x_\alpha}
\end{equation}
Assume for the time being that $x_\alpha\neq0$ for all
$\alpha\in\cA$. Clearly $dW=0$ with respect to the affine coordinates
is a stronger condition that $dW=0$ with respect to the homogeneous
coordinates. The difference between these conditions is precisely the
$(\C^*)^r$ orbits we quotient out by anyway. So we may unambiguously
refer to the condition $dW=0$ for the critical point set. If we set
any $x_\alpha$'s equal to zero, we need to be a little more careful.

If $dW=0$ has a solution where all the homogeneous coordinates
$x_\alpha$ are nonzero then there is a noncompact $\C^*$-orbit. This
yields bad theories along the primary component $\Delta_{\cB}$.

The other possibilities for the discriminant are where some of the
coordinates $x_\alpha$ are zero. This happens potentially for any face of
$\Conv\cA$ \cite{AP:genmir}. That is, if $\Xi$ is a face of $\Conv\cA$
then we set
\begin{equation}
  x_\alpha=0\quad\hbox{iff}\:\alpha\in\Xi\cap\cA\\
\end{equation}
and check if the $\C^*$ orbit from $\nu$ is compact. Setting these
$x_\alpha$'s to zero will set any monomial in $\cB$ to zero unless
that monomial contains $x_\alpha$ with exponent 0. That is, we
restrict $W$ to having monomials in $\cB\cap\Xi^\vee$. The discriminant
$\Delta_{\cB\cap\Xi^\vee}$ will therefore describe solutions of the
problem $dW=0$ for this restricted theory without the points
$\alpha\in\Xi$. That is we are solving
\begin{equation}
\frac{\partial W}{\partial x_\alpha}=0,\quad\forall\alpha\not\in\Xi.
\end{equation}
But this isn't what we want. We require {\em all\/} the derivatives of
$W$ to vanish. The condition $\Delta_{\cB\cap\Xi^\vee}=0$ is therefore
weaker than finding a noncompact $\C^*$-orbit. Note that we can also
state the weaker condition by saying that the discriminant
$\Delta_{\cB\cap\Xi^\vee}$ represents the solutions of
\begin{equation}
x_\alpha\frac{\partial W}{\partial x_\alpha}=0,\quad\forall\alpha.
\end{equation}

We see that the locus of singular theories in the B-side moduli
space is a subset of the vanishing locus of
\begin{equation}
  \prod_{\Xi\subset\Conv\cA} \Delta_{\cB\cap\Xi^\vee}. \label{eq:EX}
\end{equation}
The reflexive condition says that the faces of $\Conv\cB$ are
precisely the dual faces of $\Conv\cA$. In this case we see that
(\ref{eq:EX}) becomes $E_\cB$ ignoring the multiplicities. 

The determinant $E_\cB$ therefore does {\em not\/} represent the locus of
singular theories for the GLSM with target space $X_\Sigma$. We
discuss the differences in section \ref{ss:compact}. In particular,
the true discriminant for $X_\Sigma$ will typically lack some of the
components of $E_\cB$ and will have different multiplicities.

%%%%%%%%%%%%%%%%%%%%%%%%%%%%%%%%%%%%%%%%%%%%%%%%%%%%%%%%%%%%%%%%%%%

\section{The A-Side Discriminant}   \label{s:Aside}

\subsection{Components}  \label{ss:comps}

If mirror symmetry works, we expect to find a model of the discriminant on
the A-side associated with the pointset $\cA$ in the same way that
the pointset $\cB$ was used in the previous section.

There are two ways the theory can become singular due to deformations
of the $a_\alpha$ parameters:
\begin{enumerate}
\item Since the $a_\alpha$'s control the K\"ahler form and thus size
  of $X_\Sigma$, we can go to some limit where $X_\Sigma$ becomes
  infinitely large and thus not compact.
\item The fields $\sigma_i$ (see below) in the GLSM may become
  massless and then fluctuate over a noncompact space.
\end{enumerate}
In the first possibility we are going to the boundary of the moduli
space. In the toric model of the moduli space coming from the
secondary fan, this means we are going to toric divisors. We saw in
the previous section that it was vertices on the convex hull of $\cB$
that gave such contributions, and so now we expect points appearing as
vertices of $\Conv\cA$ to account for these parts of the discriminant.

Now let us focus on the second possibility. We refer to \cite{MP:inst}
for a description of the $\sigma_i$ fields. All we really need to know
about them here is that they have masses controlled by
$a_\alpha$'s. For generic $a_\alpha$'s these fields are massive and
can be integrated out.

We can find the location of the discriminant following \cite{MP:inst}
by setting the $\sigma_i$ fields to have large vevs. This gives masses
to the $x_\alpha$ fields which can then be integrated out. The
equations of motion then put a condition on the $a_\alpha$'s which
beautifully reproduces the Horn uniformization formula
(\ref{eq:Horn}). The $\sigma_i$'s play the r\^ole of the parameters
$\lambda_i$ in this equivalence. It is important to note
\cite{MP:inst} that in order to get the agreement with the Horn
uniformization formula, it is necessary do compute a tadpole-like
one-loop correction term when integrating out the $x_\alpha$ fields.

Anyway, we have nicely reproduced $\Delta_\cA$ for part of the
discriminant on the A-side. What about the other components? This was
described heuristically in \cite{MP:inst} and then more carefully in
\cite{AP:genmir}. For non-singular models, we give vevs to the scalar
fields $x_\alpha$ and this is viewed as a Higgs phase. When we give
vevs to the $\sigma_i$ fields, which live in vector superfields, we
consider it to be in a Coulomb phase. Naturally we can consider a
mixed Coulomb--Higgs phase where some $x_\alpha$'s have vevs and some
$\sigma_i$'s have vevs. To this end we split the GLSM into two parts
so we can put one part in the Coulomb phase and the other in the Higgs
phase. Choose a subgroup $\widehat G_h\subset\widehat G$ and let $\widehat
G_c$ be the quotient $\widehat G/\widehat G_h$, with $\pi_h$ the
quotient map. Let $e_\alpha$ be a basis for $\Z^{\cA}$. Then define
\begin{equation}
\begin{split}
\cA_h &= \{ \alpha\in\cA: \pi_h Qe_\alpha=0\}\\
\cA_c &= \cA - \cA_h.
\end{split}
\end{equation}

In \cite{AP:genmir} the following was proven:
\begin{prop} There is a commutative diagram
\begin{equation}
\xymatrix{
&0&0&0\\
0\ar[r]&M_c\ar[r]\ar[u]&\Z^{\cA_c}\ar[r]^-{Q_c}\ar[u]&
  \widehat G_c\ar[u]\ar[r]&0\\
0\ar[r]&M\ar[r]^{A^t}\ar[u]&\Z^{\cA}\ar[r]^Q\ar[u]&
  \widehat G\ar[u]_{\pi_h}\ar[r]&0\\
0\ar[r]&M_h\ar[r]\ar[u]&\Z^{\cA_h}\ar[u]\ar[r]^-{Q_h}&
   \widehat G_h\ar[u]\ar[r]&0\\
&0\ar[u]&0\ar[u]&0\ar[u]\\
}  \label{eq:split}
\end{equation}
with all rows and columns exact.
Here $Q_c$ is defined as $\pi_hQ$ restricted to $\Z^{\cA_c}$, and
$Q_h$ exists uniquely to make the diagram commute. $M_c$ and $M_h$ are
defined as the kernels of $Q_c$ and $Q_h$ respectively.
\end{prop}

The first row in (\ref{eq:split}) represents the Coulomb GLSM and the
last row represents the Higgs GLSM. Of central interest to us is this
Higgs GLSM. This GLSM is constructed from the points $\cA$ living in
quotient lattice $N_h$ of $N$ dual to the sublattice $M_h\subset
M$. While the pointset $\cA$ lives in a hyperplane, the Higgs pointset
$\cA_h$ need {\em not\/} lie in a hyperplane in $N_h$. That is, the Higgs
data need not obey the \CY\ condition.

We want consider the conformal field theory associated to the GLSM
which means we want to consider the endpoint of infrared
renormalization group flow. If the \CY\ condition is violated, the IR
flow will move us in the A-model moduli space of this Higgs model. In
\cite{AP:genmir} it was argued that we should impose the condition
$N_{c,\R}$ should meet $\Conv\cA$ along a face. If we do not obey this
condition, then the IR flow simply takes us to a model where we do
obey this condition, and we would just end up over-counting the various
possibilities.

So, for each face $\Upsilon$ of $\Conv\cA$ we can split the GLSM into
a Higgs part and Coulomb part. The data in the Coulomb part is given
by the pointset in $\Upsilon$ and we can simply redo the computation
we did at the start of this section and obtain a discriminant
component $\Delta_{\cA\cap\Upsilon}$. Comparing to (\ref{eq:Adet}) we
see we are getting close to the $E_\cA$ determinant.

\subsection{Multiplicities} \label{ss:mult}

The Coulomb part of the theory describes the decompactification and
thus the fact that the GLSM has become singular. The Higgs part
therefore describes the ``internal'' part of this singular theory. It
should therefore be a natural source of the description of the
multiplicity of the singularity. 

Let $\Upsilon$ be a face of $\Conv\cA$. The Coulomb theory is
associated to the pointset $\cA\cap\Upsilon$ and the Higgs theory is
associated to the pointset in the quotient $\cA/\Upsilon$ (where the
quotient refers to modding out by the linear subspace spanned by the
face $\Upsilon$).

Let us first ignore the superpotential and consider the noncompact
Higgs GLSM given by $\cA/\Upsilon$. The original pointset $\cA$ lay in
a hyperplane which forces $Z_\Sigma$ to be \CY. The pointset
$\cA/\Upsilon$ will, in general, not lie in a hyperplane. The IR flow
process will therefore be nontrivial. This IR flow process has been
studied in \cite{Harvey:2001wm,Morrison:2004fr}.

The general idea is that IR flow favours points closer to the origin
rather than further away, since the closer points are more relevant in
the final conformal field theory. In toric geometry language, a
concave fan represents a positive canonical class, which corresponds
to negative curvature. The IR flow is roughly given by Ricci flow
which would make negatively curved regions larger. Conversely, convex
fans correspond to positive curvature which shrink down in the IR
flow.

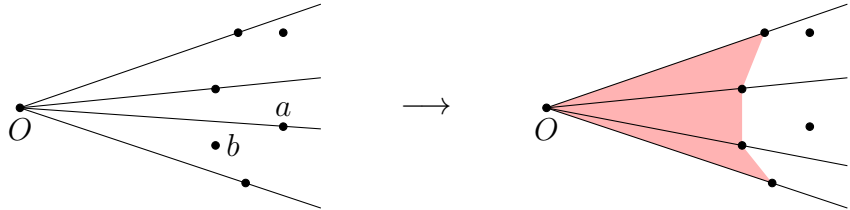
\begin{figure}
\begin{center}
\begin{tikzpicture}[scale=1.0]
\begin{scope}
\draw (-2,1) -- (2,1.4);
\draw (-2,1) -- (2,0.72);
\filldraw (-2,1) circle (0.05);
\filldraw (1,0) circle (0.05);
\filldraw (0.6,0.5) circle (0.05) node[anchor=west] {$b$};
\filldraw (0.6,1.25) circle (0.05);
\filldraw (0.9,2) circle (0.05);
\filldraw (1.5,2) circle (0.05);
\filldraw (1.5,0.75) circle (0.05) node[anchor=south] {$a$};
\draw (-2,1) -- (2,-1/3);
\draw (-2,1) -- (2,1+4/2.9);
\draw (-2,1) node[anchor=north] {$O$};
\draw (3.4,1) node {$\longrightarrow$};
\end{scope}
\begin{scope}[xshift=70mm]
\fill[fill=red!30,line width=0pt] (-2,1) -- (0.9,2) -- (0.6,1.25) 
     -- (0.6,0.5) -- (1,0) -- (-2,1);
\filldraw (-2,1) circle (0.05);
\filldraw (1,0) circle (0.05);
\filldraw (0.6,0.5) circle (0.05);
\filldraw (0.6,1.25) circle (0.05);
\filldraw (0.9,2) circle (0.05);
\filldraw (1.5,2) circle (0.05);
\filldraw (1.5,0.75) circle (0.05);
\draw (-2,1) -- (2,-1/3);
\draw (-2,1) -- (2,1+4/2.9);
\draw (-2,1) node[anchor=north] {$O$};
\draw (-2,1) -- (2,1.4);
\draw (-2,1) -- (2,0.23);
\end{scope}
\end{tikzpicture}
\end{center}
\caption{The result of IR flow.} \label{fig:flow}
\end{figure}

The result of a typical IR flow is shown in figure \ref{fig:flow}. We
begin with the fan on the left. The point $a$ represents an irrelevant
operator or region of positive curvature. This will shrink down in the
flow. The point $b$ represents a relevant operator which was
ignored. By including this in the fan, we get a region of negative
curvature that will expand. Thus we end up with the fan on the right.

The result of IR flow is that we reproduce the results of section
\ref{ss:bad}. We obtain
\begin{prop}
The points used for constructing the rays of
the fan are those appearing as the vertices of $\Conv(S/\Upsilon)_+$.
\end{prop}

We need to obtain the multiplicities from this geometry. A natural
first guess would be the Witten index, or (orbifold) Euler characteristic.
A more sophisticated guess, which will be shown to be more appropriate
in section \ref{s:mon}, would be the rank of the topological K-theory
group $K_0$ of the model. Happily, for toric varieties, these two
numbers coincide as we discuss later.

We would therefore like to compute the Euler characteristic of the fan
over $\Conv(S/\Upsilon)_+$. We claim this coincides perfectly with
$u(\Upsilon)$:
\begin{prop}
The (orbifold) Euler characteristic of the toric variety associated
with the IR limit of the Higgs GLSM is
\begin{equation}
  \chi = \Vol\Bigl(\Cone(S/\Upsilon)_+ - \Conv(S/\Upsilon)_+\Bigr),
\end{equation}
\end{prop}
  
To prove this note that a toric variety is the union of all the
various torus orbits. The Euler characteristic is additive so we just
need to add up the contribution of these orbits.

Let $\tau$ be a face of the fan. The orbit cone correspondence
associates a torus-orbit $O(\tau)$ with this face. If the torus orbit
$O(\tau)$ has positive dimension then it is a torus itself and has
$\chi=0$. So we only care about torus fixed points, i.e., maximal
cones in $\Sigma$ under the orbit-cone correspondence. Without loss of
generality we may assume the fan has been subdivided to be
simplicial. In this case, each maximal cone corresponds to an affine
toric variety of the form $\C^d/H$ for some finite abelian group
$H$. The volume of this maximal cone is given by the order of the
group, $|H|$. However, the orbifold Euler characteristic is given by
the number of twisted sectors, which is also $|H|$. Thus the total
Euler characteristic is the total volume of the fan. \QED

We have, therefore, agreement for the multiplicities of the components
of the GKZ determinant and the multiplicities associates with the Higgs
theories. But it is important to note that we have not yet imposed
the superpotential constraint.

\subsection{Compact vs Noncompact Mirror Symmetry}  \label{ss:compact}

In the previous section we computed the multiplicities for the Higgs
toric variety associated with $\cA/\Upsilon$. This is a noncompact
toric variety. This is exactly the Higgs part of theory associated
with the noncompact $Z_\Sigma$ as opposed to the compact $X_\Sigma$.

We should also note that computation of the GKZ $\cB$-determinant was
not the same as the computation of the singularities of
$X_\Sigma$. In particular, {\em logarithmic\/} differential forms were
used for computing the vanishing of $dW$ in (\ref{eq:logdW}). This
amounts to using logarithmic coordinates for the GLSM.

We therefore arrive at
\begin{prop}
  The discriminants of singular theories are consistent, including
  multiplicities, with mirror symmetry for the GLSM moduli spaces if
  we compare the K\"ahler moduli space of the {\em noncompact\/} toric
  variety $Z_\Sigma$ with the mirror theory $Y$ but using logarithmic
  coordinates.
\end{prop}
For the latter, we would write the superpotential as a polynomial in
$\exp(Y_\beta)$. This idea of noncompact \CY\ s being mirror to
compact theories with log coordinates is not new. It appeared in the
work of \cite{Hori:2000kt}, where mirror GLSM's are directly obtained by a
duality construction. Logarithmic coordinates also appeared earlier in 
\cite{MR1653024}.

The astute reader will notice that we have cheated badly by switching
to a statement about noncompact mirror symmetry. The motivation for
the appearance of singularities associated to the discriminant came
from a noncompact target space. In our picture of noncompact mirror
symmetry the targets space is now always noncompact!

The solution to this quandary is as follows. We should actually
compute some well-behaved correlation functions and see where they
diverge. However, it is fairly clear what would happen. On the A-side
we would use operators corresponding to cohomology classes with
compact support. Such compactness of support would be lost when the
$\sigma$-vacua appear if the support of the representative of the
class touched that point, i.e., if the corresponding $x_\alpha$'s
vanished. Thus we would argue that the same argument to locate the
discriminant works just as well in the noncompact as the compact case.

To pass back to the compact version of mirror symmetry we would expect to modify
multiplicities on both sides. On the B-side we need to restore the log
coordinates back to normal coordinates. The analogue of (\ref{eq:logdW})
in normal coordinates is not as easy to compute. In particular a
useful determinantal computation such as theorem 2.7 of \cite{GKZ:book}
relies on a smoothness condition which will be violated in most cases.

Actually the issues of multiplicities in the compact case is quite
awkward. There is no longer an agreement between total cohomology and
Euler characteristic. Furthermore, when we categorify, we no longer
have agreement with topological K-theory and algebraic K-theory. We
will therefore mainly restrict our attention to multiplicities in the
noncompact model in this paper.

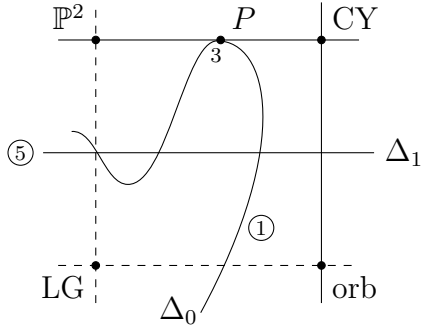
\begin{figure}
\begin{center}
\begin{tikzpicture}
\draw[dashed] (-0.5,0) -- (3.5,0);
\draw[dashed] (0,-0.5) -- (0,3.5);
\draw (-0.5,3) -- (3.5,3);
\draw (3,-0.5) -- (3,3.5);
\filldraw(0,0) circle (0.05) node[anchor=north east] {LG};
\filldraw(3,0) circle (0.05) node[anchor=north west] {orb};
\filldraw(0,3) circle (0.05) node[anchor=south east] {$\P^2$};
\filldraw(3,3) circle (0.05) node[anchor=south west] {CY};
\filldraw(1.66,3) circle (0.05) node[anchor=south west] {$P$};
\draw (-0.7,1.5) -- (3.7,1.5) node[anchor=west] {$\Delta_1$};
\begin{scope}[xshift=-55.2,yshift=137.6,scale=0.088,yscale=-1]
\draw ( 18.4 , 34.7 )
.. controls ( 22.1 , 34.4 ) and ( 23.3 , 42.7 ) ..
( 26.9 , 42.7 )
.. controls ( 32.8 , 42.8 ) and ( 35.5 , 21.0 ) ..
( 40.4 , 21.0 )
.. controls ( 48.1 , 21.0 ) and ( 51.9 , 36.2 ) ..
( 37.9 , 62.1 );
\end{scope}
\draw (1.1,-0.6) node {$\Delta_0$};
\draw (2.2,0.5) node[circle,inner sep=1pt,draw] {\scriptsize 1};
\draw (-1.0,1.5) node[circle,inner sep=1pt,draw] {\scriptsize 5};
\draw (1.6,2.8) node {\scriptsize 3};
\end{tikzpicture}
\end{center}
\caption{The discriminant for $\P^4_{\{9,6,1,1,1\}}$.}  \label{fig:18}
\end{figure}

A case where a rigorous argument shows that the multiplicity {\em
  must\/} change as we go to compact model is that of
$\P^4_{\{9,6,1,1,1\}}$ as studied in \cite{CFKM:II}. This weighted
projective space has a $\Z_3$ quotient singularity along
$[x_1,x_2,0,0,0]$. Let $Z$ be the total space of the canonical line
bundle over this projective space blown-up along this locus. Note that
$Z$ still has quotient singularities. Let $X$ be the degree 18
hypersurface in this blown-up weighted projective space. $X$
generically misses the singularities in $Z$ and is then smooth.  This
example has $h^{1,1}=2$ and the discriminant is shown in figure
\ref{fig:18}. In particular, there are two components $\Delta_0$ and
$\Delta_1$. Although it is far from obvious from the figure, there is
a symmetry of this model on $X$ which exchanges these two components.
To restore this $\Z_2$ symmetry on the moduli space, one must do a
sequence of blow-ups as performed in \cite{CFKM:II}.

At the noncompact level, $Z$ has multiplicities 1 and 5 respectively
for these components. The symmetry is therefore broken. In order to
restore the symmetry for $X$, the multiplicity for $\Delta_1$ must be
reduced to one.

\subsection{Vertices of $\Conv\cA$}   \label{ss:vertices}

At the start of section \ref{ss:comps} we said that there were two
sources of singularities in the K\"ahler moduli space. We have dealt
with the second source, namely when the $\sigma_i$'s become massless. 
These reproduce the Horn uniformization formula, which accounts for
all the factors in $E_\cA$ {\em except the contribution for vertices}.
The vertex contributions are associated to monomials in $b_\beta$ and
are thus associated to toric divisors in the moduli space. That is, we
associate these singularities to phase limits such as large radius
limits.

Recall the determinant for the quintic:
\begin{equation}
 E = b_1^4b_2^4b_3^4b_4^4b_5^4(b_0^5 + 5^5b_1b_2b_3b_4b_5).
\end{equation}
How do we explain in the multiplicity of 4 associated to the large
radius limit $b_1\to 0$? We will understand this more clearly, and
convincingly, when we describe spherical functors later but argue for
now as follows. In the noncompact model, the $\P^4$ expands to
infinite size in the large radius limit. Thus 8-branes, 6-branes,
4-branes and 2-branes wrapping this space become infinitely
massive. Suppose we wish to normalize the masses of D-branes such that
nothing associated to compact cycles becomes infinite. A natural
choice would be to normalize the mass of 8 branes, wrapping the whole
$\P^4$, to be one. If we do this, the masses of 6-branes, 4-branes,
2-branes and 0-branes are all massless in the large radius limit. It
is as if a $\P^3$'s worth of D-branes have become massless. Then
$\chi(\P^3)=4$ explains the observed multiplicity.

Note, furthermore, that the GKZ construction of this multiplicity
directly constructs $\P^3$. Namely, if $\Upsilon$ is the 0-dimensional
face of $\Conv\cA$ associated with the vertex $b_1$, the fan over
$\cA/\Upsilon$ is indeed the fan for (a line bundle over) $\P^3$.

Now consider the general case. The vertices of $\Conv\cA$ are
associated with toric divisors in $Z$. Let $D$ be the divisor
associated with the vertex point. Now assume we have a basis for 
divisors which includes $D$. That is any divisor class will be written
\begin{equation}
  sD + D_1,  \label{eq:DJ}
\end{equation}
where $D_1$ is a linear combination of the remaining basis
elements. If $Z$ were compact we could use Poincar\'e duality to map
divisors classes into $H^2(Z)$ and thus write the K\"ahler class in
this form. Actually since we are computing masses of D-branes wrapping
compact objects, we can restrict attention to compact subspaces of $Z$.
We can thus restrict attention to compactly supported $H^2$ and then
Poincar\'e duality works. We will, therefore, imagine that the
K\"ahler form is dual to (\ref{eq:DJ}). Now consider some maximal
dimensional compact toric subspace $K\subset Z$ of dimension
$d_K$. The volume of $K$ is then
\begin{equation}
  \int_KJ^{d_K} = K\cap(sD+d_1)^{d_K}.
\end{equation}
Suppose $m$ is such that $K\cap D^m\neq0$ and $K\cap D^{m+1}=0$. Then,
in the large $s$ limit, the volume of $K$ goes as $s^m$.

The volume of $K\cap D$ is given by $K\cap D\cap(sD+d_1)^{d_K-1}$. The
volume of this clearly goes as $s^{m-1}$. Thus, if we normalize the
D-branes wrapping all of $K$ to remain finite mass, the masses on
$D\cap K$ go to zero. In this sense, $D$ represents the massless
D-branes.

Note that this picture is not perfect. While D-branes on the divisor
$D$ go massless in the large $s$ limit, there can be D-branes going
massless on other divisors in this limit too. We give a more precise
meaning to what the relationship between $D$ and massless D-branes is
in section \ref{ss:Horja}.

%%%%%%%%%%%%%%%%%%%%%%%%%%%%%%%%%%%%%%%%%%%%%%%%%%%%%%%%%%%%%%%%%%%

\section{D-Brane Monodromy}   \label{s:mon}

We can get much more information about the discriminant by looking at
the D-brane category. In particular we will consider the automorphisms
of the category of D-branes induced by monodromy around the components
of the discriminant. We view the multiplicity as the rank of the
K-theory of a category associated with the monodromy. In this sense we
are {\em categorifying\/} the discriminant multiplicity. We will need
to use the technology of derived categories and triangulated
categories for this section as it by far the best way to understand
monodromy. We refer to \cite{me:TASI-D}, for example, for a review of
the basic ideas.

\subsection{Spherical Functors} \label{ss:spher}

Let $\mathsf{D}$ be the D-brane category, which will either be the
bounded derived category of coherent sheaves (in a geometric phase) or
the Fukaya category. For B-type branes, the category will be either
$\DC(Z_\Sigma)$ or $\DC(X_\Sigma)$ depending on whether we use the
noncompact or compact version of the GLSM.

Consider an exact functor
\begin{equation}
F:\mathsf{A} \to \mathsf{D},
\end{equation}
with a right adjoint $R$ and a left adjoint $L$. We can then define
two more functors.\footnote{To be pedantic, we need to go to
  dg-categories to truly define functors but this is of no practical
  consequence for this paper \cite{AL:ST}.} The {\bf cotwist} $C$
and {\bf twist} $T$ associated to $F$ are the cones on the unit and
counit of the adjunction
\begin{align*}
C &= \Cone(1 \xrightarrow\eta RF) & T &= \Cone(FR \xrightarrow\epsilon 1). 
\end{align*}
The functor $F$ is called {\bf spherical} if any two of the following
conditions are true,
\begin{enumerate}
\item
$T$ is an autoequivalence of $\mathsf{D}$,
\item $C$ is an autoequivalence of $\mathsf{A}$,
\item $R\cong LT[-1]$,
\item $R\cong CL$,
\end{enumerate}
in which case all 4 conditions are true \cite{AL:ST}.

If $\mathsf{A}$ is a triangulated category then a 
{\bf Serre functor} $S_{\mathsf{A}}$ is an autoequivalence of
$\mathsf{A}$ such that, within $\mathsf{A}$,
\begin{equation}
  \Hom(\mathsf{a},\mathsf{b}) = 
        \Hom(\mathsf{b},S_{\mathsf{A}}\mathsf{b})^\vee.
\end{equation}
If $\mathsf{A}$ and $\mathsf{D}$ both admit Serre functors then
condition 4 above is equivalent to $S_\mathbf{D}FC\cong FS_\mathbf{A}$.

The spherical twist is quite natural in the context of D-brane
monodromy as we now explain. As described in \cite{AD:Dstab}, D-brane
monodromy is induced by changes in stability conditions as one goes
around a loop in the moduli space of complexified K\"ahler forms.
Recall that stable D-branes $\mathsf{c}$ have a central charge
$Z(\mathsf{c})$ and a grade $\xi(\mathsf{c})$ such that
\begin{equation}
  \xi(\mathsf{c}) = -\frac1{\pi}\arg(Z(\mathsf{c}))\pmod{2}.
\end{equation}
In a distinguished triangle
\begin{equation}
\xymatrix@C=5mm{\mathsf{a}\ar[rr]&&\mathsf{b}\ar[dl]\\
&\mathsf{c}\ar[ul]|{[1]}},
\end{equation}
$\mathsf{c}$ is stable with respect to decay into $\mathsf{a}[1]$ and
$\mathsf{b}$ if and only if $\xi(\mathsf{b})-\xi(\mathsf{a})<1$.

Let us suppose $\mathsf{a}$ is a D-brane which can become
massless. Assume, for simplicity of argument, there is a simple zero
in $Z(\mathsf{a})$ at some point $P$ in the moduli space. Suppose also
that $\mathsf{b}$ is massive at $P$, i.e., $Z(\mathsf{b})\neq0$. Now
go counterclockwise around a small loop enclosing
$P$. $\xi(\mathsf{b})$ will not change but $\xi(\mathsf{a})$ will
decrease, and the result may change $\mathsf{c}$ from being unstable
to stable.

So, in going around the loop, we can change the set of stable
objects. The monodromy on $\mathsf{D}$ reflects this change in
stability.  In its most na\"\i ve form, the idea in monodromy is that,
given a starting D-brane, $\mathsf{b}$, you allow all possible
massless D-branes $\mathsf{a}$ to bind to $\mathsf{b}$ as above to
turn $\mathsf{b}$ into $\mathsf{c}$. The autoequivalence induced by
the monodromy then replaces $\mathsf{b}$ by this new bound state.

This means that monodromy around $P$ is associated to a map $M$ from
objects in $\DC(X)$ to {\em massless\/} objects in $\DC(X)$, such that
the action of monodromy replaces $\mathsf{b}$ by 
\begin{equation}
\Cone(M(\mathsf{b})\to\mathsf{b}). \label{eq:c1}
\end{equation}

Speaking loosely for a moment, let $\mathsf{A}$ be a category whose
objects are the D-branes massless at $P$. The map $M$ must then factor
through $\mathsf{A}$ and we write $M=FR$:
\begin{equation}
\xymatrix@1{\mathsf{D}\ar[r]^R&\mathsf{A}\ar[r]^F&\mathsf{D}.}
  \label{eq:Mfactor}
\end{equation}
At present we have only defined these maps as set maps on objects in
the categories. The natural thing to do is to elevate them to
functors. Why we should do this depends on an understanding of the
morphisms in $\mathsf{A}$. We will not attempt this here but will
assume these maps are functors from now on.

The map in (\ref{eq:c1}) therefore gives a map of functors
\begin{equation}
\xymatrix@1{FR\ar[r]&\id},
\end{equation}
which, presumably, we want to be a natural map. This is instantly
recognizable as the counit of adjunction and so we say $R$ is the
right-adjoint of the functor $F$. Thus we arrive at monodromy being
given by the spherical functor associated with $\mathsf{A}$ being
the category of massless D-branes.

While we have motivated associating a spherical functor with
monodromy, we cannot really claim to have derived it since we really
don't say why $F$ and $R$ should be functors rather than just
maps on objects. Having said that, if we do associate monodromy with a spherical
functor, it does seem that $\mathsf{A}$ gives the category of massless
D-branes.

Actually even this latter statement requires some care. There is an
alternative picture for how the monodromy turns $\mathsf{b}$ into
$\mathsf{c}$. Suppose that the added $\mathsf{a}$ is simply a direct sum
$\mathbf{b}\oplus\mathsf{c}[-1]$.  Then, rather than saying
$\mathsf{c}$ is a equivalent to a massless D-brane $\mathsf{a}$ added
to $\mathsf{b}$, we have obtained $\mathsf{c}$ crudely from
$\mathsf{b}$ by simply annihilating it completely and then adding
$\mathsf{c}$. While the central charge of
$\mathbf{b}\oplus\mathsf{c}[-1]$ is zero at $P$, we don't consider it as a
massless D-brane as it is not stable.

One can take this to an extreme. In \cite{Seg:allS} it was shown by
Segal that {\em any\/} autoequivalence of the D-brane category can be
written as a spherical twist. This construction was done by taking
{\em all\/} elements of $\mathsf{A}$ to be of the form
$\mathbf{b}\oplus\mathsf{c}[-1]$ in the above sense.

Given an autoequivalence there is thus an ambiguity in specifying the
source category $\mathsf{A}$. Clearly we want to find the one
generated by stable massless objects and thus be ``as far as possible'' from
the Segal construction. This appears to be satisfied in the cases we
consider below as in each case $\mathsf{A}$ is generated by stable
branes we know to be massless.

%%%%%%

\subsection{Generic Monodromy}

Given that we associate a spherical twist with $\mathsf{A}$, the
category of massless D-branes, and that the Higgs GLSM describes the
singularity structure, we arrive at an obvious conjecture:
\begin{conjecture} \label{conj:D}
The generic monodromy around a component of the discriminant is given
by a spherical twist, where $\mathsf{A}$ is the category of D-branes in
the associated Higgs GLSM.
\end{conjecture}

This conjecture can be applied in the noncompact case $Z_\Sigma$ as
well as the compact case $X_\Sigma$. In the compact case we impose
the superpotential constraint. In derived category language this means
that we look at the category of matrix factorizations of the
superpotential within the Higgs GLSM.

For the primary component of the discriminant we know that the Higgs
GLSM is trivial, i.e., that of a point, and thus the category
$\mathsf{A}$ is generated by a single object. We therefore expect
generic monodromy around this component to be associated with a single
D-brane becoming massless. That is, the monodromy is a Seidel--Thomas
autoequivalence \cite{ST:braid}. This is a conjecture apparently due
to Kontsevich \cite{Kont:mon}.

Note that our conjecture does not fully specify the spherical
twist. We do not specify the functor $F:\mathsf{A}\to\DC(X)$, we
merely specify what $\mathsf{A}$ is. This is because we have not given
enough information to specify $F$. The monodromy requires
a choice of basepoint and a particular path around the discriminant.
A change in path can result in a quite different functor $F$. We will
see this very explicitly in section \ref{sss:ambig}.

\subsection{Relationship to a Horja Twist}   \label{ss:Horja}

There is clearly a relationship between the above conjecture and the
spherical twists due to Horja \cite{Horj:EZ}.  Recall that
given\footnote{We've already used the letter $Z$, so we replace $Z$ in
  Horja's notation by $U$,}
\begin{equation}
\xymatrix{ E \ar@{^{(}->}[r]^i\ar[d]^q&Z\\ U},  \label{eq:EZ}
\end{equation}
there is an associated spherical functor
\begin{equation}
  i_*q^*: \DC(U) \to \DC(Z). 
\end{equation}
That is, $\DC(U)$ represents the category of massless D-branes
$\mathsf{A}$..  The general idea is that something equivalent to
$\DC(U)$ becomes massless as $E$ is collapsed onto $U$ \cite{AKH:m0}.

Now consider the following process associated to a face $\Gamma$ of
$\Conv\cA$. Begin with a fan $\Sigma$ given by a triangulation of
$\cA$, and the associated toric variety $Z_\Sigma$. Now do a minimal
detriangulation of $\cA$, to a polytopal (not necessarily simplicial)
decomposition, so that the face $\Gamma$ is not subdivided. That is,
$\Gamma$ is now the face of a polytope . This produces a toric variety
we will call $Z_{\Sigma\backslash\Gamma}$. The detriangulation
produces a blow-down
\begin{equation}
  r:Z_\Sigma\to Z_{\Sigma\backslash\Gamma}.
\end{equation}
$\Gamma$ now spans a cone $\sigma_\Gamma$ in the fan, and there is an
associated sub-toric-variety $V(\sigma_\Gamma)$ given as an orbit
closure. (See section 3.2 of \cite{CLS:ToricVar}.) So we have an
inclusion
$j:V(\sigma_\Gamma)\hookrightarrow Z_{\Sigma\backslash\Gamma}$. But
$V(\sigma_\Gamma)$ is exactly the target space of the (noncompact)
Higgs GLSM. So we have a diagram
\begin{equation}
\xymatrix{ E_\Gamma \ar@{^{(}->}[r]^{\hat\imath}\ar[d]^{\hat
    q}&Z_\Sigma\\ V(\sigma_\Gamma)}, \label{eq:EZhat}
\end{equation}
where $E_\Gamma$ is $r^{-1}jV(\sigma_\Gamma)$.

We have identified the massless D-brane category as coming from the
Higgs GLSM and thus we identify it with $D(V(\sigma_\Gamma))$.  So we
identify $V(\sigma_\Gamma)$ with $U$ in Horja's picture.  So,
identifying the spherical functor with $i_*q^*$, we have cast the
desired monodromy in the form of a Horja twist.

The difficulty in comparing our conjecture to the Horja twist is when
we try to establish exactly which path in the moduli space generates
this automorphism, but we can describe a simple case as follows. 

The Horja picture is usually applied to the case of a codimension one
wall in the secondary fan between two phases.  The way this
description is extracted from the toric data is explained in detail in
section 2.6 of \cite{AdAs:masscat}. This wall is associated to a
$\P^1$ in the moduli space, and to a ``circuit'' in the pointset in
sense of chapter 7 of \cite{GKZ:book}.\footnote{Briefly, a circuit is
  a set of points satisfying precisely one affine relation involving
  all the points.}  The Horja twist is then associated to monodromy
around the unique point in the intersection of this $\P^1$ with the
discriminant that is not one of the two
limit points. This is called the ``wall monodromy''.

The following proposition is clear:
\begin{prop}
If the points of $\cA\cap\Gamma$ form a circuit then the corresponding wall
monodromy coincides with the generic monodromy around
$\Delta_{\cA\cap\Gamma}$.
\end{prop}
As this is probably well-known, we sketch the idea briefly. (We will
also see an example to illustrate the idea in section
\ref{sss:octic2}). The homogeneous coordinates $b_\beta$ in the
circuit can be combined to form a unique invariant affine coordinate
$x$ on the moduli space. This affine coordinate is also an affine
coordinate on the $\P^1$ in the moduli space. Furthermore, the
function $\Delta_{\cA\cap\Gamma}$ can be written purely (up to an
overall factor) as a function of $x$. Thus this component of the
discriminant intersects the $\P^1$ transversely. This proves the
proposition.

Given that the Horja twist can be proven to give the wall monodromy
\cite{HHP:linphase}, we see that our conjecture is true in this case.

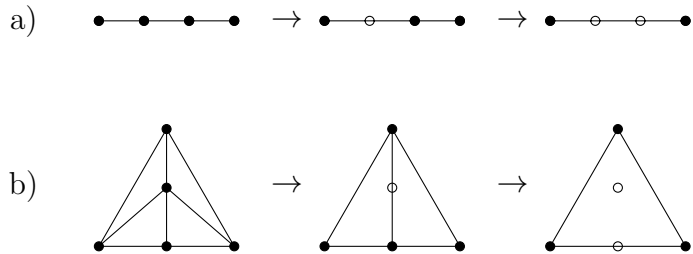
\begin{figure}
\begin{center}
\begin{tikzpicture}
\begin{scope}[scale=0.6]
\draw (0,0) circle(0.1) -- (1,0) circle(0.1) -- (2,0) circle(0.1)
  -- (3,0) circle(0.1);
\filldraw (0,0) circle (0.1);
\filldraw (1,0) circle (0.1);
\filldraw (2,0) circle (0.1);
\filldraw (3,0) circle (0.1);
\end{scope}
\draw(2.5,0) node {$\to$};
\begin{scope}[shift={(3,0)},scale=0.6]
\draw (0,0) circle(0.1) -- (1,0) circle(0.1) -- (2,0) circle(0.1)
  -- (3,0) circle(0.1);
\filldraw (0,0) circle (0.1);
\filldraw (2,0) circle (0.1);
\filldraw (3,0) circle (0.1);
\end{scope}
\draw(5.5,0) node {$\to$};
\begin{scope}[shift={(6,0)},scale=0.6]
\draw (0,0) circle(0.1) -- (1,0) circle(0.1) -- (2,0) circle(0.1)
  -- (3,0) circle(0.1);
\filldraw (0,0) circle (0.1);
\filldraw (3,0) circle (0.1);
\end{scope}
\begin{scope}[shift={(0,-3)},scale=0.6]
\draw (0,0) -- (3,0) -- (1.5,2.6) -- (0,0);
\draw (1.5,0) -- (1.5,2.6);
\draw (0,0) -- (1.5,1.3) -- (3,0);
\filldraw(0,0) circle (0.1);
\filldraw(1.5,0) circle (0.1);
\filldraw(3,0) circle (0.1);
\filldraw(1.5,1.3) circle (0.1);
\filldraw(1.5,2.6) circle (0.1);
\end{scope}
\draw(2.5,-2.2) node {$\to$};
\begin{scope}[shift={(3,-3)},scale=0.6]
\draw (0,0) -- (3,0) -- (1.5,2.6) -- (0,0);
\draw (1.5,0) -- (1.5,2.6);
\filldraw(0,0) circle (0.1);
\filldraw(1.5,0) circle (0.1);
\filldraw(3,0) circle (0.1);
\draw(1.5,1.3) circle (0.1);
\filldraw(1.5,2.6) circle (0.1);
\end{scope}
\draw(5.5,-2.2) node {$\to$};
\begin{scope}[shift={(6,-3)},scale=0.6]
\draw (0,0) -- (3,0) -- (1.5,2.6) -- (0,0);
\filldraw(0,0) circle (0.1);
\draw(1.5,0) circle (0.1);
\filldraw(3,0) circle (0.1);
\draw(1.5,1.3) circle (0.1);
\filldraw(1.5,2.6) circle (0.1);
\end{scope}
\draw(-1,0) node {a)};
\draw(-1,-2.2) node {b)};
\end{tikzpicture}
\end{center}
\caption{Two possibilities for non-circuits.}  \label{fig:2poss}
\end{figure}

Many complications arise when there are more points in $\cA\cap\Gamma$ than
needed to form a circuit. Let us consider two ways in this can
happen as shown in figure \ref{fig:2poss}. These figures show a
non-circuit. The first arrow shows a change in triangulation as one crosses a
codimension one face in the secondary fan. The last figure on each row
corresponds to the detriangulation $\Sigma\backslash\Gamma$.

The simplest case would be $\cA$ consisting of 4 points in a line as
shown in figure \ref{fig:2poss}a. A fan over this gives the resolution
of the quotient singularity $\C^2/\Z_3$. The exceptional set here is
two $\P^1$'s. This example is associated to an $\SU(3)$
Seiberg--Witten theory.  Crossing the wall to a neighboring phase
blows down one of the $\P^1$'s whereas the detriangulation of this set
blows down {\em both\/} $\P^1$'s. The wall monodromy corresponds to
the generic monodromy around the discriminant and thus corresponds to
blowing down only a single $\P^1$. The detriangulation of the face,
blowing both $\P^1$'s down, should more naturally be associated with
the Argyres--Douglas point \cite{DA:SU3}. In this case it is clear
that the generic monodromy around the component of the discriminant is
{\em not\/} given na\"\i vely by the Horja twist associated to a
detriangulation of the face.

Both maps in the case of \ref{fig:2poss}b correspond to a
fibration. The first map plays the role of $q$ in (\ref{eq:EZ}) for
wall monodromy and the two maps composed plays the role of $\hat q$ in
(\ref{eq:EZhat}) for the generic monodromy. So we have a diagram
\begin{equation}
\xymatrix{&E\ar@{^{(}->}[r]^{i}\ar[d]^{q}&Z_\Sigma\\ 
F\ar[r]&U\ar[d]\\
&V(\sigma_\Gamma)
} \label{eq:EUV}
\end{equation}
We claim that this corresponds to a non-transverse collision between
the $\P^1$ joining the limit points and the discriminant. Rather than
develop this idea in general, we will just give some examples below.
In particular we will highlight the case when $U\to V(\sigma_\Gamma)$
is a projective space fibration.

In general the situation comparing wall monodromy to generic monodromy
is a combination of the effects exemplified by the two cases in figure
\ref{fig:2poss}. Furthermore, the factored map $E\to U\to
V(\sigma_\Gamma)$ need not be a fibration. We won't consider these
cases here.

We can also look at the components of the discriminant associated to
the vertices of $\Conv\cA$. In this case, no blowing down at all is
necessary --- $V(\sigma_\Gamma)$ is already a divisor in $Z$. The
associated Horja twist is the correct one. The spherical twist
corresponding an inclusion of a divisor $D$ corresponds \cite{AL:ST}
to tensoring by a line bundle $\O(D)$, consistent with a large radius
limit monodromy. The questions arises, however, exactly why in this
case the Horja monodromy is associated to a limit point, and not a
point in the wall between phases. The exact rules are not clear to us.
 
Rather than try to establish the exact rules, let us present some
interesting examples. These serve to show the relationship between our
conjecture and the Horja twists along phase walls. We can also view
these examples as proving the conjecture true in these cases. We also
use these examples to demonstrate the issue of going from the
noncompact $Z$ to the compact $X$.

\subsubsection{The Octic}  \label{sss:octic2}

We refer back to figure \ref{fig:octic} for the geometry of the
discriminant in the case of the octic. The large radius \CY\ phase has
two neighbouring phases, the orbifold and the $\P^1$ hybrid. We first
consider Horja's wall monodromy in these two cases.

First we take the hybrid phase. In the compact picture we have the
entire \CY\ threefold $X$ collapsing down to $\P^1$. But to compare to
our conjecture we need the ambient noncompact version of the
collapse. The weighted projective space $\P^4_{\{2,2,2,1,1\}}$ is singular
along a $\P^2$. We can blow this singular set up to get a smooth
compact 4-fold $S$.  $S$ is a $\P^3$-bundle over $\P^1$. In the \CY\
phase, the space $Z$ is the canonical line bundle over $S$. Going to
the hybrid phase collapses $S$ to the base $\P^1$. Thus the phase
changing collapse $q:E\to U$ is given by $q:S\to\P^1$. The category of
massless D-branes for the singularity in the wall between the phases
is therefore $\DC(\P^1)$.

This wall monodromy is not generic monodromy however. It is
well-known how to disentangle the generic monodromy from the wall
monodromy. For explicit details see \cite{me:navi,AKH:m0}. For a
spherical functor with source category $\mathsf{A}$, let the
corresponding spherical twist be denoted $T_{\mathsf{A}}$. If $L$ is a
line bundle then let $L$ also denote the automorphism $-\otimes L$. By
\cite{Bei:res} we have $\DC(\P^1) = \langle\O,\O(1)\rangle$. Then, by
Horja, the wall monodromy is
\begin{equation}
\begin{split}
T_{\DC(\P^1)} &= T_{\langle\O,\O(1)\rangle},\\
&= T_{\langle\O\rangle}\circ T_{\langle\O(1)\rangle},
\end{split}
\end{equation}
but from the link topology at this point on the discriminant
\cite{me:navi}, we have
\begin{equation}
\begin{split}
T_{\DC(\P^1)} &= (T_{\mathsf{A}} \circ \O(1))^2 \circ \O(1)^{-2}\\
& = T_{\mathsf{A}} \circ \O(1) \circ  T_{\mathsf{A}} \circ
\O(1)^{-1}\\
&= T_{\mathsf{A}} \circ T_{\mathsf{A}\otimes \O(1)},
\end{split}
\end{equation}
where $\mathsf{A}$ is the massless D-brane category for meridians
around a generic point on $\Delta_0$, the primary component of the
discriminant. So we can identify
\begin{equation}
  T_{\mathsf{A}} = T_{\langle\O\rangle}.
\end{equation}
We are therefore consistent with the above conjecture. The massless
D-brane for this primary component is a single object. In this case
$i_*q^*\O\cong\O_S$.  This wall monodromy associated component of the
discriminant is actually an example of the type shown in figure
\ref{fig:2poss}b and equation (\ref{eq:EUV}). The component of the
discriminant in this case is the primary component $\Delta_0$ and thus
we expect $V(\sigma_\Gamma)$ to be simply a point. $U$ is given by
$\P^1$. We claim the fact that $U$ is a $\P^1$-bundle over
$V(\sigma_\Gamma)$ gives rise to the link topology above. We will
discuss this again in section \ref{sss:3p} when $V(\sigma_\Gamma)$ is
not trivial.

Restricting to the compact octic hypersurface $X$, we see that the
massless D-brane for the primary component of the discriminant is
$\O_X$.

Now consider the component $\Delta_1$ in figure \ref{fig:octD}. The is
associated to the bottom edge of $\Conv\cA$ in figure \ref{fig:octic}
with 3 points. This is a circuit and we are in the situation described
above. Thus $\Delta_1$ intersects the $\P^1$ connecting the \CY\ and
orbifold phase transversely. The wall monodromy between these two
phases is the same thing as the generic monodromy around this
component of the discriminant as discussed above. Using the methods of
\cite{AdAs:masscat}, we see that $E$ is associated to the exceptional
divisor of the blow-up. This is a noncompact space which is the total
space of the line bundle $\O(-4,0)$ over $\P^2\times\P^1$. $Y$ is the
then the blow-down of this to $\O_{\P_2}(-4)$.  Horja then tells us
the category of massless D-branes is $\DC(\O_{\P_2}(-4))$. This is
entirely consistent with our conjecture as expected. Quotienting out
the space spanned by this edge leaves us with the GLSM Higgs data for
$\O_{\P_2}(-4)$.

Since $\O_{\P_2}(-4)$ deformation retracts to $\P^2$ we have an Euler
characteristic of 3. The derived category is similarly of K-theory
rank 3. Thus we are consistent with the multiplicity of 3 in the
discriminant in figure \ref{fig:octD}.

Note that it is not obvious how to pass to the compact model
$X_\Sigma$ in this case. $Y$ should be viewed now as a genus 3 curve
in $\P^2$. That is, the category of massless D-branes should be the
derived category of this higher-genus curve. The rank of topological
K-theory is 2 and so one might predict that the multiplicity of the
discriminant should drop to 2 when passing to the compact model.

The component of the discriminant passing through the \CY\ and
orbifold phase is associated to the vertex $b_1$ (or $b_2$ or
$b_3$). This yields the divisor $V(\sigma_{b_1})$ given by $\O(-8)$ over
$\P^3_{2,2,1,1}$. The space is singular so we cannot use the Euler
characteristic directly. However, the K-theory is of rank 6,
consistent with the figure.

Finally the component of the discriminant passing through the \CY\ and
$\P^1$ limit is associated with $\O_{\P^3}(-4)$, which is consistent
with the multiplicity of 4.

Again the multiplicities of these latter two components will
presumably drop when we pass to the compact model.

\subsubsection{An example with an ambiguity}   \label{sss:ambig}

\begin{figure}
\begin{center}
\begin{tikzpicture}
\draw[dashed] (-0.5,0) -- (3.5,0);
\draw (0,-0.5) -- (0,3.5);
\draw (-0.5,3) -- (3.5,3);
\draw (3,-0.5) -- (3,3.5);
\filldraw(0,0) circle (0.05) node[anchor=north east] {$\P^1$};
\filldraw(3,0) circle (0.05) node[anchor=north west] {$\P^3$};
\filldraw(0,3) circle (0.05) node[anchor=south east] {Exo};
\filldraw(3,3) circle (0.05) node[anchor=south west] {CY};
\begin{scope}[xshift=-67.5,yshift=150,scale=0.088,yscale=-1]                   ]
\draw ( 33.7 , 54.9 ) ..
controls ( 20.6 , 27.5 ) and ( 56.6 , 51.2 )
.. ( 20.8 , 47.8 ) ..
controls ( 3.3 , 45.9 ) and ( 11.2 , 12.5 )
.. ( 26.7 , 13.1 ) ..
controls ( 39.5 , 12.0 ) and ( 48.8 , 31.6 )
.. ( 40.0 , 39.2 ) ..
controls ( 48.2 , 32.3 ) and ( 60.8 , 39.0 )
.. ( 61.0 , 44.5 ) ..
controls ( 61.1 , 56.7 ) and ( 42.7 , 43.7 )
.. ( 42.9 , 66.9 );
\end{scope}
\draw (1.1,-0.6) node {$\Delta_0$};
\draw (2.1,1.0) node[circle,inner sep=1pt,draw] {\scriptsize 1};
\draw (-0.3,2.3) node[circle,inner sep=1pt,draw] {\scriptsize 3};
\draw (3.3,2.3) node[circle,inner sep=1pt,draw] {\scriptsize 6};
\draw (0.7,2.7) node[circle,inner sep=1pt,draw] {\scriptsize 4};
\draw (3.2,1.3) node {\scriptsize 3};
\draw (2.8,1.71) circle [x radius=.07, y radius=.04, rotate=30];
\draw[->] (3.4,1.71) node[anchor=west] {$\O_X$} -- (2.9,1.71);
\draw (1.3,3.2) circle [x radius=.07, y radius=.04, rotate=30];
\draw[->] (1.9,3.4) node[anchor=west] {$\O_E$} -- (1.4,3.2);
\end{tikzpicture}
\end{center}
\caption{The discriminant for the exoflop example.}  \label{fig:exoD}
\end{figure}
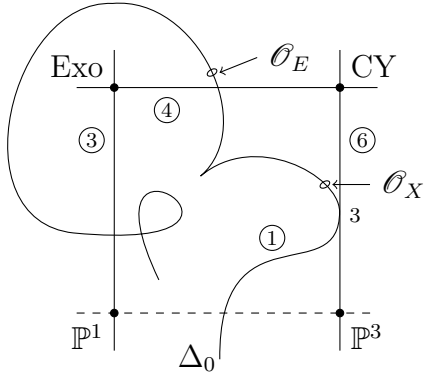

Here we give an example where a single component of the discriminant
is associated with two apparently different massless D-branes. We use
an example where the \CY\ phase consists of $Z_\Sigma$ being the
canonical line bundle of the $\P^1$-bundle $\P(\O(-1)\oplus\O)$ over
$\P^3$. This was studied in \cite{AdAs:masscat}, for example, and has
charge matrix
\begin{equation}
  Q=\begin{pmatrix}-3&-1&0&1&1&1&1\\-2&1&1&0&0&0&0\end{pmatrix}.
             \label{eq:Q6}
\end{equation}
The discriminant is of the form
\begin{multline}
E_\cA = b_1^3b_2^4b_3^6b_4^6b_5^6b_6^6(256b_1^5b_2^4 -
3^35^5b_2b_3^2b_4^2b_5^2b_6^2 + 18000b_0b_1^2b_2^2b_3b_4b_5b_6
\\- 256b_0^2b_1^4b_2^3 + 3800b_0^3b_1b_2b_3b_4b_5b_6
+96b_0^4b_1^3b_2^2 - 27b_0^5b_3b_4b_5b_6
-16 b_0^6b_1^2b_2 + b_0^8b_1).
\end{multline}
The moduli space and discriminant are depicted in figure
\ref{fig:exoD}. Note that, beside the vertex monomials associated with
limit points, there is only one component, $\Delta_0$, of the
discriminant. This component has a quite rich geometry, however, having
both a node and a cusp.

The CY$-\P^3$ transition is very similar to the octic above. The Horja
picture has $E$ given by the $\P^1$-bundle over $\P^3$ collapsing to
the base $Y\cong\P^3$. The category of massless D-branes associated to
this wall monodromy is thus given by
$\DC(\P^3)=\langle\O_{\P^3}, \O_{\P^3}(1), \O_{\P^3}(2), \O_{\P^3}(3)
\rangle$.
However, $\Delta_0$ intersects the line joining these limit points
with multiplicity 4. After disentangling the wall monodromy from the
generic monodromy around $\Delta$ we find the generic monodromy
associated to a single massless D-brane $\O_{\P^3}$. In terms of the
noncompact ambient space $Z$, this massless D-brane is $q^*\O_{\P^3}$,
which is the structure sheaf of the whole $\P^1$-bundle over $\P^3$.
That is, the support of the sheaf representing this massless D-brane
has complex dimension 4.  Restricting to the compact $X$, this
massless D-brane is $\O_X$, of complex dimension 3.

Now consider the CY--Exoflop transition. Now the $\P^3$ base collapses
to a point. The Horja wall monodromy is a simple Seidel--Thomas
autoequivalence associated to the massless D-brane $\O_W$, where
$W\cong\P^3$ is given by $x_0=x_1=0$. Since $\Delta_0$ intersects this
wall transversely, this is also the generic monodromy around
$\Delta_0$. Thus the monodromy is associated with a single massless
D-brane of the form $W\cong\P^3$. Restricting to the compact $X$, we
see a massless D-brane corresponding to a cubic surface, dP$_6$, in $W$.

We have therefore arrived at two different answers for the D-brane
massless category associated with the generic monodromy around
$\Delta_0$. In compact language we either get $\O_X$ or $\O_E$, where
$E$ is the del Pezzo surface dP$_6$. We show this in figure
\ref{fig:exoD}. This does not break our main conjecture, however. In
both cases the massless D-brane category is generated by a single
object. It is trivially the same intrinsic category in each case. What
differs is how the spherical functor $F$ maps this category into
$\DC(X)$ or $\DC(Z)$.

What this shows is that the combinatorial data for the GKZ determinant
can tell us what the category $\mathsf{A}$ is, but we require more
data to specify the functor $F$.

\subsubsection{A Three Parameter Example}   \label{sss:3p}

It would be nice see an example of (\ref{eq:EUV}) where $U$ and
$V(\sigma_\Gamma)$ are distinct and nontrivial. To see this we need to
go to a 3-parameter example. We will consider the case where $X$ is (a
resolution of) a degree 24 hypersurface in
$\P^4_{\{12,8,2,1,1\}}$. $X$ is also then an elliptic fibration over
the Hirzebruch surface $\HS_2$. The matrix of charges is
\begin{equation} Q = 
\begin{pmatrix}
-6 & 3 & 2 & 0 & 0 & 0 & 0 & 1 \\
0 & 0 & 0 & 1 & 0 & 0 & 1 & -2 \\
0 & 0 & 0 & 0 & 1 & 1 & -2 & 0
\end{pmatrix}
\end{equation}
The model has 8 phases and the secondary polytope is isomorphic to a cube.

Three phases are adjacent to the \CY\ phase:
\begin{itemize}
\item A hybrid model given by a \LG-fibration over $\HS_2$.
\item A K3-fibration over $\P^1$. The fibres are singular and the
  resulting space has a curve of $\Z_2$-quotient singularities.
\item An elliptic fibration over $\P^2_{\{2,1,1\}}$. This also has a
  curve of $\Z_2$-quotient singularities.
\end{itemize}

Computing the GKZ determinant we find
\begin{equation}
  E_{\cA} = 2^83^{12}b_1^{12}b_2^{16}b_3^{12}b_4^{12}b_5^{12}
      \Delta_0\Delta_1^5\Delta_2^6.  \label{eq:3pD}
\end{equation}
The cone in the secondary fan corresponding to the \CY\ phase gives an
affine toric patch in the moduli space with coordinates $x=b_0^{-6}b_1^3b_2^2b_7$,
$y=b_3b_6b_7^{-2}$ and $z=b_4b_5b_6^{-2}$. We then have
\begin{equation}
\begin{split}
  \Delta_0& =b_0^{24}(2^{22}3^{12}x^4y^2z - 2^{20}3^{12}x^4y^2
       +\ldots+1728x - 1)\\
  \Delta_1&= b_7^4(64y^2z-16y^2+8y-1)\\
  \Delta_2&=b_6^2(4z-1)
\end{split}
\end{equation}

\begin{figure}
\begin{center}
\begin{tikzpicture}
\draw (-3,-1) -- (0,0) -- (3,-1);
\draw (0,0) -- (0,-3);
\filldraw(0,0) circle (0.05) node[anchor=south] {Smooth};
\filldraw(-3,-1) circle (0.05) node[anchor=east] {$\HS_2$ hybrid};
\filldraw(3,-1) circle (0.05) node[anchor=west] {K3 fibration};
\filldraw(0,-3) circle (0.05) node[anchor=north] {Elliptic fibration};
\draw (-2,-2) .. controls (-1.8,-0.2) .. (-1,-1.6);
\draw (-2.3,-1.4) node {$\Delta_0$};
\draw (-1,-1.1) node {\scriptsize pt};
\draw (-1.8,-0.4) node {\scriptsize 4};
\draw (1,-1.6) .. controls (1.8,-0.2) .. (2,-2);
\draw (2.3,-1.4) node {$\Delta_1$};
\draw (0.8,-1.1) node {$\scriptstyle\P^1_{\{3,2\}}$};
\draw (1.8,-0.4) node {\scriptsize 2};
\draw (-1,-2.4) -- (1,-2.6);
\draw (-1.3,-2.4) node {$\Delta_2$};
\draw (0.8,-2.3) node {$\scriptstyle\P^2_{\{3,2,1\}}$};
\end{tikzpicture}
\end{center}
\caption{The discriminant for $\P^4_{\{12,8,2,1,1\}}$.}  \label{fig:24}
\end{figure}
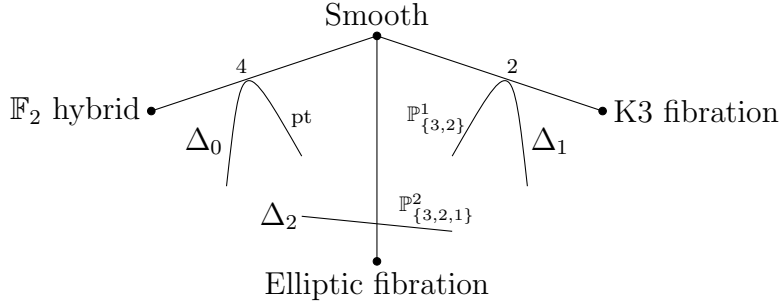

We show how the components of the discriminant hit the lines between
phase limits in figure \ref{fig:24}. Note this is a very schematic
drawing --- the discriminant in this case is complex dimension 2. We
will consider the wall monodromies for all 3 cases in this figure.

$\Delta_2$ has a transverse collision with the line joining the \CY\ phase to the 
elliptic fibration over $\P^2_{\{2,1,1\}}$. The Horja wall monodromy
corresponds to a collapse in the ambient $Z$ of (line bundles over)
$\P^2_{\{3,2,1\}}\times\P^1$ to $\P^2_{\{3,2,1\}}$. The rank of
K-theory of $\P^2_{\{3,2,1\}}$ is 6 consistent with the exponent of
$\Delta_2$ in (\ref{eq:3pD}). Again we saw a similar effect in the
octic. In this case $U\cong V(\sigma_\Gamma)\cong\O_{\P^2_{\{3,2,1\}}}(-6)$.

Now consider $\Delta_1$. This component hits with degree 2 the line
joining the \CY\ phase to the K3-fibration. By setting $x=0$, we get
the same geometry for the collision as we did in the case of the
octic's $\Delta_0$. The homotopy of the link complement is thus the
same as in the case of the octic.

According to Horja's picture, the wall monodromy is associated with a
collapse of the subspace, $E\subset Z$, given by $x_7=0$, to the
subspace $U$ in the neighboring phase given by $x_3=x_6=0$. In
particular we find the tower of vertical maps in (\ref{eq:EUV})
becomes
\begin{equation}
\xymatrix{
E \cong \O_{\P^1_{\{3,2\}}}(-6)\times \HS_2\ar[d]\\
U \cong \O_{\P^1_{\{3,2\}}}(-6)\times \P^1\ar[d]\\
**{!<8mm,0mm>}{V(\sigma_\Gamma) \cong \O_{\P^1_{\{3,2\}}}(-6)}}
\end{equation}

The best way to understand this wall monodromy is from a theorem due
to Orlov \cite{MR1208153}. Suppose we have a vector bundle $E\to M$ of
rank $r$. There is then an associated bundle $\P(E)$ with fibre given
by a projective space $\P^{r-1}$. Furthermore, there is a canonically
defined sheaf $\O(1)$ on $\P(E)$ (see, for example, section II.7 of
\cite{Hartshorne:}). Orlov's theorem then tells us that there is a
semiorthogonal decomposition
\begin{equation}
  \DC(\P(E)) = \langle \DC(M), \DC(M)\otimes\O(1), \DC(M)\otimes\O(2),
    \ldots, \DC(M)\otimes\O(r-1)\rangle.
\end{equation}
This allows us to interpret the connection with generic monodromy the
same way as we did for the octic. We put $U=\P(E)$ and
$M=V(\sigma_\Gamma)$ and $r=2$ and obtain
\begin{equation}
  \DC(U) = \langle \DC(V(\sigma_\Gamma)),
  \DC(V(\sigma_\Gamma))\otimes\O(1)\rangle.
\end{equation}
 The massless D-brane category for the wall
monodromy is $\DC(U)$ but, because of the degree 2 intersection, we
relate this to a generic monodromy around this component of the
discriminant to $V(\sigma_\Gamma)$. Thus we are in complete agreement
with our main conjecture again.

It is not hard to see how this construction always works in the case
that $U$ is of the form of the projectivization $\P(E)$ of a bundle over
$V(\sigma_\Gamma)$. 

$\Delta_0$ hits the line joining the \CY\ phase and the hybrid at a
point of degree 4. In this case $U\cong\HS_2$ and $V(\sigma_\Gamma)$ is a
point according to the conjecture. In order to verify the consistency
of this assertion we would need to analyze the structure of the
homotopy of the complement of the discriminant near the collision
point. This is difficult and we do not attempt it. What we would
expect, however, is that we would build the derived category of
$\HS_2$ from an exceptional collection four point-like objects.

%%%%%%%%%%%%%%%%%%%%%%%%%%%%

\section{Discussion} \label{s:conc}

We have seen that the GKZ A-determinant plays a key role in the gauged
linear $\sigma$-model. Obviously it would be very nice to have a
direct GLSM computation of the GKZ determinant. One would expect this
to come from a suitable computation of a partition function. However,
the GKZ determinant has an anomalous degree as given in section
\ref{sss:anomaly} so some care would be needed in this approach. It
would be very satisfying to see a direct computation of this anomaly
too from
the field theory.

By indirectly analyzing the GKZ determinant in terms of singularities
and monodromy,
we appear to have a nice picture of the discriminant for the
noncompact story. Each component of the discriminant is associated to
a category of massless D-branes which describes, via a spherical
twist, the generic monodromy in the D-brane category. An explicit form
of the functor in this spherical twist requires more information, such
as a choice of large radius limit phase and a choice of path from this
limit to the meridian where we do the monodromy. But the intrinsic
structure of the massless D-brane category requires no such
choice. This category encodes the ``badness'' of the singularity.

We have proven that the multiplicities match between the GKZ picture
and the GLSM picture and we have a natural conjecture for the
monodromy.  Before we can prove the latter, we need to understand the
meaning of the morphisms in this massless category. Obviously they
play an essential r\^ole but we don't know any motivation for their
interpretation yet. It would be nice to analyze the Higgs GLSM
directly and see if there is an interpretation of the morphisms in
terms of open strings. Presumably, methods such as \cite{HHP:linphase}
may be of help here.

In the noncompact case we have a clear picture of the multiplicities
of the components of the discriminant. One can easily extract the rank
of the algebraic K-theory group $K_0$. In the examples in this paper,
the massless D-brane category had an exceptional collection and this
rank is simply the number of elements in this collection. The niceties
of toric geometry imply that this rank is also the rank of topological
$K_0$, which encodes the rank of the group of massless D-brane
charges, and the Witten index.

The compact case is much harder. The massless D-brane category could
easily now be something like a K3 category, where the Chow ring is
infinitely generated, and thus the algebraic K-theory group is
unpleasant. Now if, one supposes that the desired multiplicity comes
from the topological K-theory one needs to extract this information
from the massless category. This can be done \cite{Blanc:topK} but it
is far from straight-forward.
  
In this paper we have avoided going down the standard route of solving
Picard--Fuchs equations and looking at periods, couplings,
etc. Presumably the meaning of multiplicity has some meaning using these
methods and one could link the results in this paper to such techniques.

%%%%%%%%%%%%%%%%%%%%%%%%%%%%%%%%%%%%%%%%%%%%%%%%%%%%%%%%%%%%%%%%%%%
\section*{Acknowledgments}

We thank Nick Addington and Ed Segal for helpful discussions. PSA is
supported by NSF grants DMS--1207708. MRP is supported by NSF grant
PHY--1521053.  Any opinions, findings, and conclusions or
recommendations expressed in this material are those of the authors
and do not necessarily reflect the views of the National Science
Foundation.

%\bibliographystyle{my-phys}
%\bibliography{string0,new}

\end{document}